\documentclass[conference,10pt]{IEEEtran}

\IEEEoverridecommandlockouts

\usepackage{setspace}
\usepackage{enumerate}
\usepackage{cite}
\usepackage{times}
\usepackage{url}
\usepackage{graphicx}
\usepackage{subfigure}
\usepackage{amsmath}
\usepackage{color}
\usepackage[ruled,figure,linesnumbered]{algorithm2e}
\usepackage{multirow}
\usepackage{array}
\usepackage{wrapfig}

\newcommand{\REM}[1]{}

\begin{document}
\title{Dynamic Load Balancing Strategies for Graph Applications on GPUs}
\author{}
\author{
\IEEEauthorblockN{
$^1$Ananya Raval,
$^2$Rupesh Nasre,
$^1$Vivek Kumar,
$^1$Vasudevan R,
$^1$Sathish Vadhiyar,
$^3$,$^4$Keshav Pingali
}
\IEEEauthorblockA{
$^1$Department of Computational and Data Sciences, Indian Institute of Science, Bangalore, India \\
$^2$Department of Computer Science and Engineering, Indian Institute of Technology, Madras, India \\
$^3$Institute for Computational Engineering and Sciences, University of Texas at Austin, USA \\
$^4$Department of Computer Science, University of Texas at Austin, USA \\
ananya.raval@gmail.com, rupesh@cse.iitm.ac.in, vivekkumar1987@gmail.com, \\vasudevan@ssl.serc.iisc.in, vss@serc.iisc.in, pingali@cs.utexas.edu
}
}
\maketitle

\begin{abstract}
Acceleration of graph applications on GPUs has found large interest due to the ubiquitous use of graph processing in various domains. The inherent \textit{irregularity} in graph applications leads to several challenges for parallelization. A key challenge, which we address in this paper, is that of load-imbalance. If the work-assignment to threads uses node-based graph partitioning, it can result in skewed task-distribution, leading to poor load-balance. In contrast, if the work-assignment uses edge-based graph partitioning, the load-balancing is better, but the memory requirement is relatively higher. This makes it unsuitable for large graphs. In this work, we propose three techniques for improved load-balancing of graph applications on GPUs. Each technique brings in unique advantages, and a user may have to employ a specific technique based on the requirement. Using Breadth First Search and Single Source Shortest Paths as our processing kernels, we illustrate the effectiveness of each of the proposed techniques in comparison to the existing node-based and edge-based mechanisms.
\end{abstract}

\section{Introduction}
\label{intro}
Large graphs are becoming ubiquitous due to the rise of large connected real life networks including social, transportation, web and gene expression networks, which can be abstracted and modeled as graphs. There is large interest in the acceleration of different graph processing algorithms and applications including breadth first search (BFS), single-source shortest path (SSSP), minimum spanning tree (MST) and betweenness centrality algorithms on GPUs~\cite{nasre-morphgpus-ppopp2013, merrill-scalablegputraversal-ppopp2012, sariyuce-bc-gpgpu2013, gharaibeh-graphsgpus-ipdps2013, ediger-graphct-tpds2013, buluc-parallelBFS-sc2011,mclaughlin-bc-sc2014} due to both the large processing needs of the applications and also due to the benchmark efforts like Graph500~\cite{graph500-web}.

Traditionally, graph applications are considered to be difficult to analyze and parallelize. This difficulty stems from the unpredictable data-access and control-flow patterns, termed as \textit{irregularity}, which is inherent in graph applications. Thus, it is quite challenging to statically predict the access pattern without any knowledge about the input graph. Therefore, most of the effective techniques towards the analysis and the parallelization of graph algorithms are dynamic in nature.

One of the key challenges while dealing with irregular graph algorithms is maintaining load-balance across GPU threads. Depending upon the graph, the task-distribution can get skewed. As an example,
many of the GPU-based implementations~\cite{nasre-morphgpus-ppopp2013, nasre-datavstoplogy-ipdps2013, merrill-scalablegputraversal-ppopp2012, sariyuce-bc-gpgpu2013, gharaibeh-graphsgpus-ipdps2013} employ a node-based task-distribution, i.e., they assign a thread to a set of graph nodes. 
Such a node-based distribution works well with the compressed sparse-row (CSR) storage format often used to store graphs compactly
(CSR format represents a graph in adjacency list format, but the adjacencies of each node are concatenated together to form a monolithic adjacency list 
of size equal to the number of edges. 
Each node's adjacency list then starts at various offsets in this monolithic list).
However, this node-based distribution is unsuitable for graphs with wide variance in the node degrees, such as the social networks. 
This is because in many graph applications, the activity at each node deals with propagating some information to its neighbors or collecting it from them.
Therefore, the work done is proportional to the degree of a node.
Thus, the node-based task-distribution can lead to high load-imbalance for skewed degree graphs. 

To address the load-balancing problem, previous research has explored edge-based task-distribution~\cite{sariyuce-bc-gpgpu2013}.
In this method, a thread is assigned a set of edges (instead of nodes as in the node-based method above).
This leads to near-perfect load balancing across threads, since each thread processes (almost) the same number of edges.
However, there are a couple of issues with edge-based processing.
First, it may not always be feasible to convert a node-based processing algorithm into an edge-based processing algorithm.
Theoretically, such a conversion requires the node activity to be \textit{distributive} which need not necessarily hold.
Second, it poses restrictions on the graph format: to assign an edge to a thread, the graph should either be in a coordinate list (COO) format or it should be converted to such a format
(COO format represents a graph as a sequence of edges with each edge as a tuple $<src, dst, wt>$).
The former consumes more memory while the latter has conversion overheads.
The memory requirement is a key factor for GPUs as they continue to have low memories (upto 12 GB).
In fact, in our experiments with large graphs, we found that the edge-based methods which rely on COO format cannot be executed due to insufficient memory.
These restrictions make edge-based task-distribution unsuitable as a general solution.

Despite the aforementioned issues, node-based and edge-based task-distribution methods are attractive because of their simplicity.
The simplicity stems from the one-time assignment of graph elements to threads.
Thus, existing methods are \textit{static} in nature.
Unfortunately, the characteristics of irregular graph algorithms often change dynamically.
The processing workload at different parts of the graph varies as the algorithm progresses.
Therefore, application of static load-balancing techniques is often inadequate, and
we need dynamic load balancing mechanisms while dealing with graph algorithms.

In this paper, we propose three techniques to address the issues with the existing methods.
The first method, which combines node-based and edge-based methods is \textit{workload decomposition}. 
It assigns a set of edges to a thread similar to the edge-based task-distribution.
However, the considered edges belong only to the set of \textit{active} nodes (i.e., nodes in the worklist where there is work to be done) resembling a node-based task-distribution.
The second method, which we call as \textit{node splitting}, avoids load-imbalance by changing the underlying graph structure.
As the name suggests, it splits a high-degree node into several small-degree nodes -- thereby reducing the load-imbalance.
The third method uses a \textit{hierarchical processing} and employs a hierarchy of worklists.
The imbalance in the task-distribution from the main (super) worklist is handled by creating several sub-worklists distributed evenly across threads and changing the number of threads proportional to the size of the sub-worklist.

Following are the primary contributions of our paper.
\begin{enumerate}
\item We identify the limitations of the currently prevalent node and edge-based approaches towards load-balancing graph algorithms on GPUs.
\item We propose three methods to address the above limitations: \textit{workload decomposition}, \textit{node splitting} and \textit{hierarchical processing}. We discuss various trade-offs associated with these methods; a method needs to be chosen depending upon the application requirements and the nature of input graphs. We argue that a single load-balancing strategy is unlikely to be suitable in all scenarios.
\item We perform a comprehensive evaluation of the existing and the proposed methods using a range of real-world and synthetic graphs and two graph algorithms: breadth first search processing and single source shortest paths computation. We illustrate the utility of each method and quantitatively show the trade-offs involved.
\end{enumerate}

The rest of the paper is organized as follows.
In Section~\ref{motivation} we explain existing node-based and edge-based task-distributions and discuss their advantages and limitations.
In Section~\ref{strategies} we propose our load balancing strategies and discuss the trade-offs involved.
In Section~\ref{exp_res} we illustrate the effectiveness of our proposed techniques using two graph algorithms.
In Section~\ref{related} we compare our work with and contrast it against other proposed methods dealing with graph applications on GPUs. 
Finally, in Section~\ref{con_fut} we conclude and mention our plans for future work.

\section{Motivation}\label{motivation}
In this section we motivate the need for dynamic load-balancing strategies.
Towards this goal, we first explain the existing static approaches, namely, node-based task-distribution and edge-based task-distribution,
along with their advantages and disadvantages.

\subsection{Node-based Task-distribution}
In node-based task-distribution, the unit of processing is a node.
Thus, when a GPU kernel is invoked, the number of threads in the launch-configuration is proportional to the number of graph nodes.
In an extreme case, one may assign each node to a different thread.
For a node-based distribution it is natural to represent the graph in a CSR format.
Such a representation is also space-efficient.
Each thread then operates on each of the assigned set of nodes, and may propagate a computed information along the set of edges incident on the node.
Therefore, the amount of work done by a thread becomes proportional to the degree of the node on which it operates.
Since this degree is unknown statically and varies across inputs, node-based task-distribution (and in general, a static load-balancing technique)
incurs a high load-imbalance if the variance in the node-degrees is large (e.g., in social networks wherein degrees follow a power-law distribution).

\begin {figure}
\centering
\subfigure[USA Road Network]{
  \includegraphics[width=0.45\linewidth]{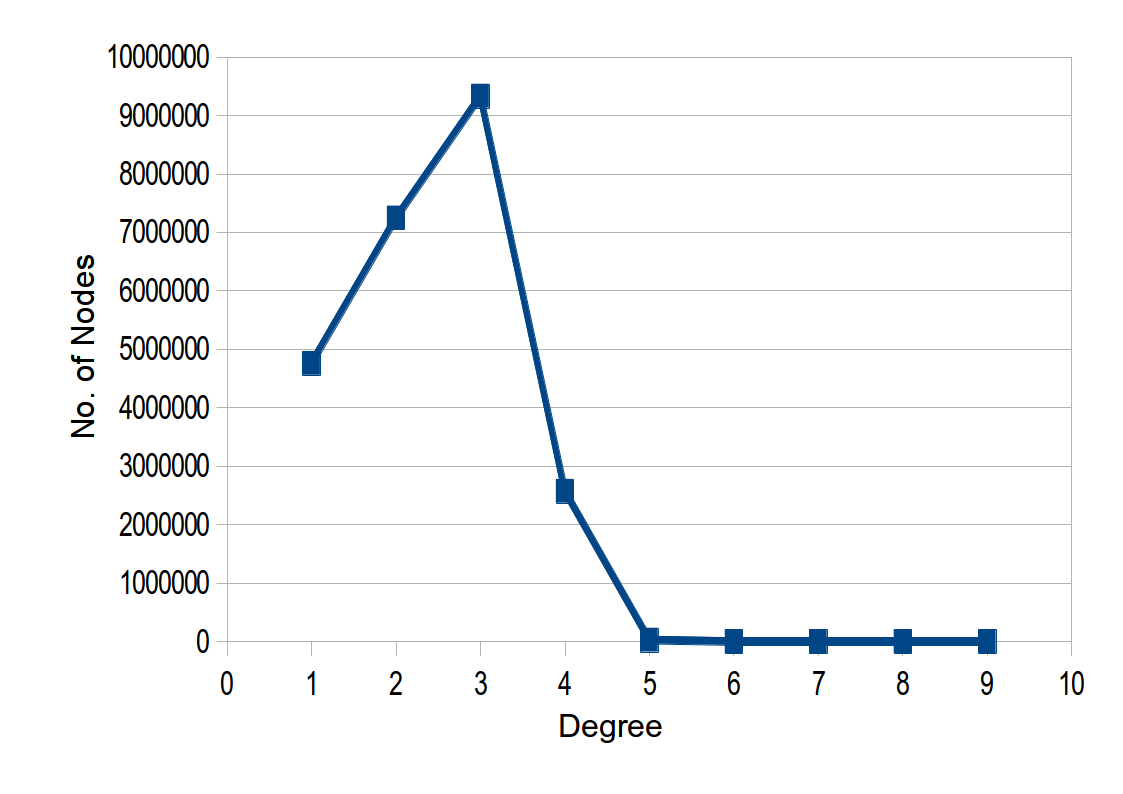}
  \label{fig:rmatdd}
}
\subfigure[Stanford Web Graph]{
  \includegraphics[width=0.45\linewidth]{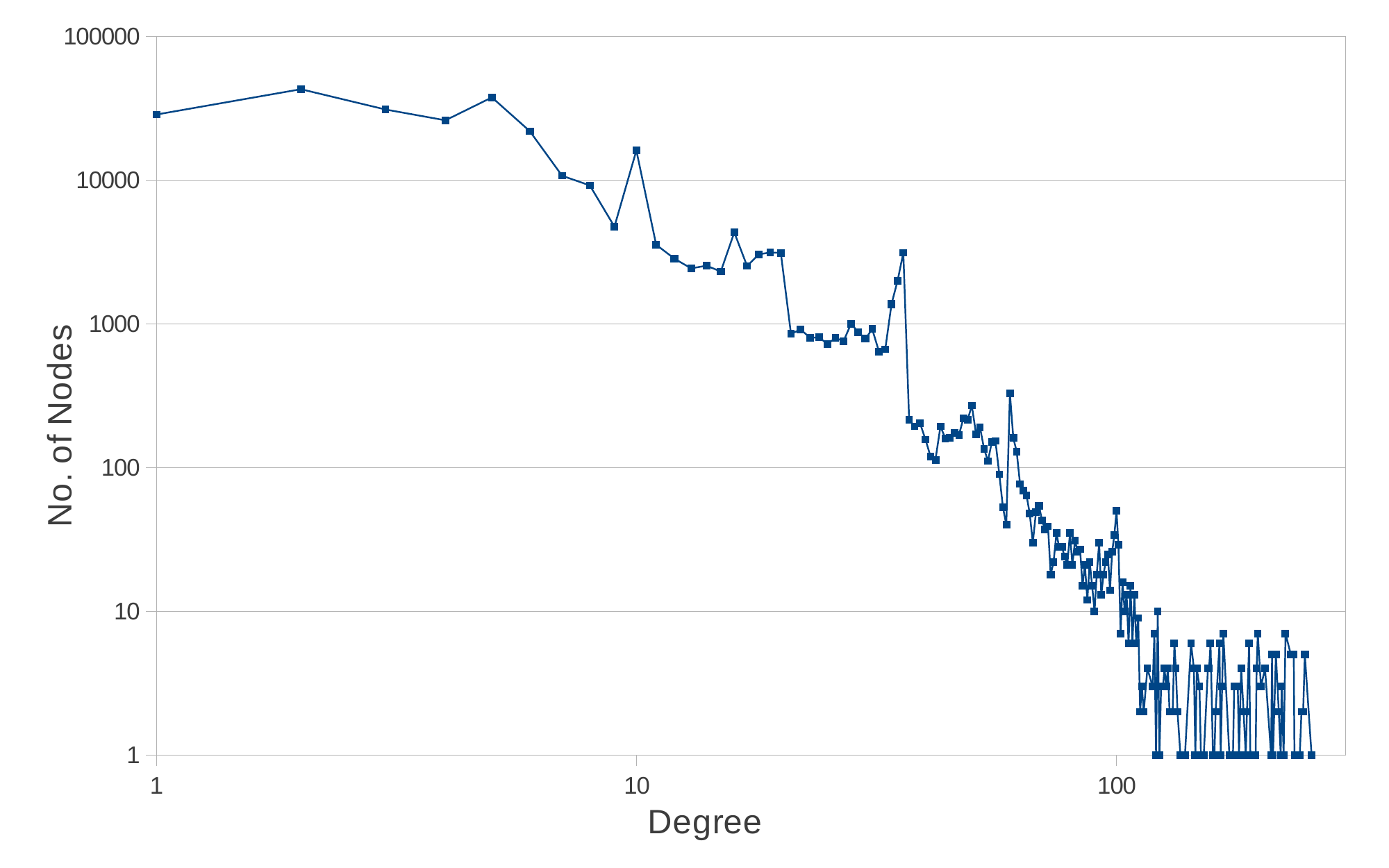}
  \label{fig:websdd}
}
\caption{Degree Distributions of Graphs and Load Imbalance}
\label{fig:dd}
\end{figure}

Figures~\ref{fig:websdd} and \ref{fig:rmatdd} show the outdegree distribution of the USA road network and Stanford web graph respectively. 
We find that the Stanford graph has a relatively larger variation in the node outdegrees (USA: minimum=1, maximum=9, average=2.4, Stanford: minimum=1, maximum=255, average=8.2).
We observe similar skewed degree distribution across other social networks as well (flickr, citations, twitter, etc.).
For such graphs, if a thread is assigned to a node, the threads operating on high outdegree nodes would dominate the computation time.
This causes the threads operating on low outdegree nodes to incur a long wait resulting in inefficient GPU resource utilization. 
In summary, node-based task-distribution, which is used in several recent works~\cite{nasre-morphgpus-ppopp2013, nasre-datavstoplogy-ipdps2013, merrill-scalablegputraversal-ppopp2012, sariyuce-bc-gpgpu2013, gharaibeh-graphsgpus-ipdps2013}, may deliver poor performance on high-skew graphs (it should be noted that load-balancing was not their primary goal).

\subsection{Edge-based Task-distribution}
In order to improve load-balance, former research has proposed edge-based task-distribution for graph algorithms on GPUs~\cite{sariyuce-bc-gpgpu2013}.
In this approach, threads operate on the graph edges;
that is, the number of threads in the kernel launch configuration is proportional to the number of graph edges.
Edge-based task-distribution provides near-perfect load-balance for propagation algorithms (such as BFS, SSSP, etc.),
since the edges can be almost evenly distributed across threads.
Such an edge-based graph partitioning is also oblivious to the degree distribution, and hence is likely to work well across different types of graphs.
Given $E$ edges and $T$ threads with $T < E$, the threads are assigned to the edges in a round-robin manner. This ensures coalesced access since the neighboring edges are assigned to consecutive threads. For our work, we spawned the GPU kernel with the maximum number of active threads possible for a given CUDA architecture.

Figure~\ref{EP-cpu-pseudocode} shows the pseudocode for edge-based parallelism using SSSP as an example.  SSSP maintains two worklists: \textit{inputWl} and \textit{outputWl} for reading and writing, respectively.  The \textit{while} loop at Lines~\ref{whilestart-line}--\ref{whileend-line} processes all the edges in the current input worklist by repeatedly calling \textit{sssp\_kernel} on the GPU. In the end, the computed distance values (attribute \textit{dist}) are copied from GPU to CPU.

Parallel edge-based SSSP requires synchronization across threads.
This is required when two edges being operated by two threads point to the same node -- resulting in a conflict.
We use atomic primitives (\textit{atomicMin}) to update the distance values.
Populating a shared worklist such as \textit{outputWl} also necessitates synchronization and we rely on atomic primitives for inserting edges.
A na\"ive way to insert each outgoing edge of a node results in a considerable overhead.
We employ work-chunking~\cite{nasre-datavstoplogy-ipdps2013} to club together multiple edges and use a single atomic instruction for adding all the edges of a node.

\begin{algorithm}[t]
\begin{small}
\SetKwInOut{Input}{input}
\SetKwInOut{Output}{output}
\SetKwProg{Fn}{}{ \{}{\}}

\Input{a graph $graph(N,E)$ with $N$ nodes and $E$ edges}

\BlankLine
$graph$.init()\\
$\forall n: dist[n] = \infty$	\label{distinit-line} \\ 
$dist[source] = 0$ 		\label{distinitsource-line} \\
$inputWl$.push($source.edges$)\\

\BlankLine
\While{$inputWl.size() > 0$} {	\label{whilestart-line}
   \textbf{\textit{sssp\_kernel}}($graph, dist, inputWl, outputWl$) \\
   $inputWl = outputWl$\\
   $outputWl$.clear() \\
}	\label{whileend-line}
copy $dist$ to CPU 

\BlankLine
\Fn(){\textbf{sssp\_kernel}($graph, dist, inputWl, outputWl$)}	{


\For{each edge $e$ assigned to me}{ \tcc{round-robin edge assignment}
  $altdist = dist[e.source] + e.weight$ \\
 
  \If{$dist[e.destination] > altdist$}{
    $dist[e.destination]$ = $altdist$ \\
    $outputWl.push(e.destination.edges)$ \label{work-chunk-line} \\
  }
}
}
\caption{SSSP Pseudocode illustrating Edge-Based Parallelism}
\label{EP-cpu-pseudocode}
\label{EP-gpu-pseudocode}
\end{small}
\end{algorithm}

\REM {
\begin{algorithm}
\begin{small}
\SetKwProg{Fn}{}{ \{}{\}}

\Fn(){\textbf{sssp\_kernel}($graph, dist, inputWl, outputWl$)}{


\For{each edge $e$ assigned to me}{
  $altdist = dist[e.source] + e.weight$ \\
 
  \If{$dist[e.destination] > altdist$}{
    $dist[e.destination]$ = $altdist$ \\
    $outputWl.push(e.destination.edges)$ \label{work-chunk-line} \\
  }
}
}
\caption{Example GPU Pseudocode for Edge-Based Parallelism (SSSP)}
\label{EP-gpu-pseudocode}
\end{small}
\end{algorithm}
}

Despite its advantages over the node-based approach, edge-based task-distribution suffers from the following limitations.
First, it may not always be possible to use an edge-based distribution.
For a node-based computation to be converted into a functionally equivalent edge-based computation, the computation needs to have a distributivity property.
A function $f$ distributes over another function $g$ if $f(g(a, b), c) \equiv g(f(a, c), f(b, c))$.
For example, in BFS, computing the minimum level for the destination node distributes over the addition (one plus) operation on the source node level.

Second disadvantage of edge-based distribution is its higher space complexity. 
A space-efficient CSR representation is unsuitable for edge-based distribution since the source node information is not duplicated across all the outgoing edges of that node.
Therefore, edge-based distribution often demands a less-normalized COO format wherein source node is duplicated across edges emanating from that source.
For a graph with $E$ edges, the COO representation requires $2E$ elements for storing the two arrays (in contrast to $N + E$ for CSR representation in case of node-based distribution).
This additional storage cost can be substantial for denser graphs, and is especially relevant for GPUs which continue to have smaller main memories (maximum 12 GB for NVIDIA GPUs).
If the vertices in the arrays are represented by 4-byte integers, the maximum number of edges that can accommodated in a system with GPU device memory of $4GB$ is 500 million for non-weighted graphs and about 350 million for weighted graphs. 
Third, edge-based task-distribution can lead to the size explosion of worklists. 
The size of the worklist can become greater than the number of edges $E$ since the outgoing edges of a node may be added redundantly to the worklist by multiple threads. 
This would require condensing the worklist and removing redundancy at the end of each GPU kernel invocation, resulting in condensing overhead~\cite{merrill-scalablegputraversal-ppopp2012}.
Large worklists also stress the memory resources.

\section{Proposed Load Balancing Strategies}
\label{strategies}
We propose three load balancing strategies: \textit{workload decomposition}, \textit{node splitting} and \textit{hierarchical processing} which we discuss in the following subsections.  
All our strategies implement data-driven GPU executions~\cite{nasre-datavstoplogy-ipdps2013} in which only the \textit{active} elements are processed using a worklist.

\subsection{Workload Decomposition}\label{workload-decomposition}
\begin{figure}
\centering
\includegraphics[scale=0.5]{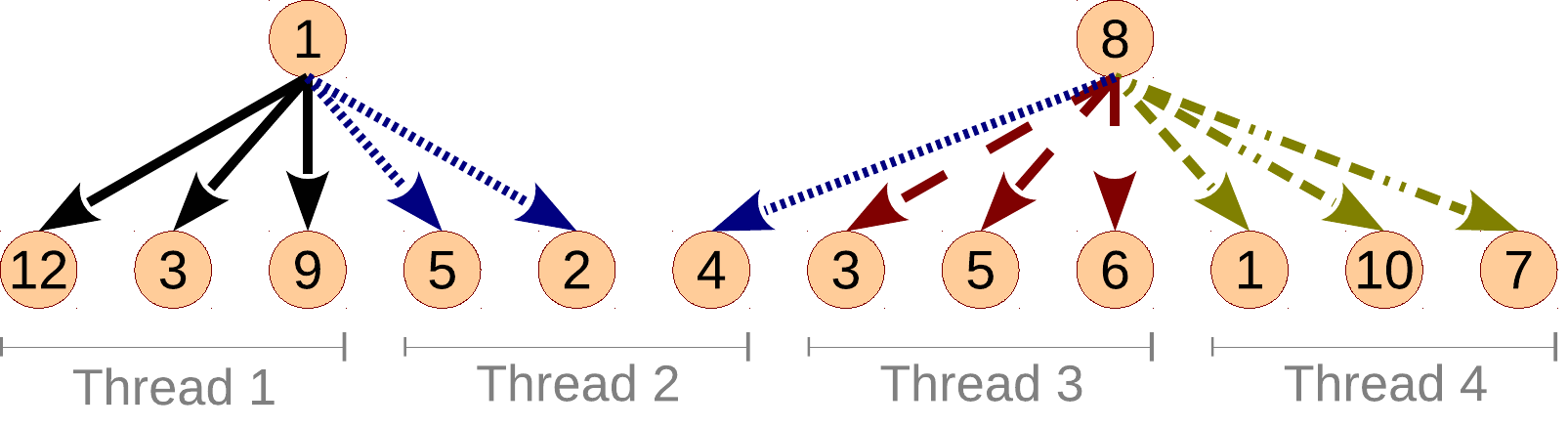}
\caption{Workload Decomposition}
\label{wd-illustration}
\end{figure}

Workload decomposition combines node-based and edge-based task-distribution.
It can be viewed as a form of space decomposition.
In this approach, the processing elements in the worklist continue to be the nodes, but the workload of the nodes, namely, the edges, are decomposed across threads using a block distribution.
$E$ number of graph edges are partitioned across $T$ threads such that each thread receives a contiguous chunk of $E/T$ edges for processing.
Thus, a given thread processes a subset of edges corresponding to a subset of nodes and all the edges outgoing from a node may not be processed by the same thread.
Figure~\ref{wd-illustration} illustrates the workload decomposition strategy with two nodes. 
In this figure, four threads process three edges each from two nodes with the first node having five outgoing edges and the second node having seven edges. 
We find that Thread 2 processes two edges from node 1 and one edge from node 8.

Figure~\ref{WD-cpu-pseudocode} presents the pseudocode for the strategy. 
Each GPU thread processes \textit{edgesPerThread} number of edges starting from a particular outgoing edge of a particular node. 
Each worklist maintains the nodes to be processed and each node's outdegree as two associative arrays.
Both the arrays are populated while updating the worklist, but only the second array (node outdegrees) is used in the prefix sum computation to determine \textit{edgePerThread} and to populate the \textit{offset} array of structure.
These offsets are calculated in the GPU in a separate kernel function \textbf{\textit{find\_offsets}} (Lines~\ref{getoffset-line-1} and \ref{getoffset-line-2}). 
A GPU thread in this kernel uses the prefix sums of the outdegrees of the nodes and the \textit{edgesPerThread} to find the particular node and edge that it has to start with for processing. 
The prefix sums of the outdegrees are also calculated on the GPU using a separate kernel. 
We use the NVIDIA's Thrust API for inclusive scan for this purpose (Line~\ref{prefixsum-line}). 
The \textit{while} loop in the kernel (Line~\ref{while-line}) handles the situation when a thread, after processing a subset of edges for a node, moves to processing the next set of edges from the next node in the worklist.

\begin{algorithm}[t]
\begin{small}
\SetKwInOut{Input}{input}
\SetKwInOut{Output}{output}
\SetKwProg{Fn}{}{ \{}{\}}

\Input{a graph \textit{graph(N,E)} with \textit{N} nodes and \textit{E} edges}
\label{offset-line} \Input{\textit{offset[numThreads]} - array of structure containing two fields: (i) \textit{NodeOffset} is the starting offset of a node in the input worklist to be processed by a thread, (ii) \textit{EdgeOffset} is the starting edge of the node to be processed by the thread}

\BlankLine
\textit{graph}.init()\\
\textit{$\forall$ n: dist[n] = $\infty$} \\
\textit{dist[source] = 0} \\
\textit{inputWl}.push(\textit{source}) \\

\textit{offsets} = \textbf{\textit{find\_offsets}}(\textit{graph, inputWl, edgesPerThread, numThreads}) \label{getoffset-line-1} \\

\BlankLine
\While{inputWl.size() $>$ 0} {
   \textbf{\textit{sssp\_kernel}}(\textit{graph, dist, inputWl, outputWl, offsets}) \\
   \textit{inputWl = outputWl} \\ 
   \textit{outputWl}.clear() \\
   \BlankLine

   \textit{prefixsum = scan(\textit{graph})}	\label{prefixsum-line}	\tcc{Use Thrust API}
   \textit{edgesPerThread = ceil(prefixsum.size() / outputWl.size())} \\
   \textit{offsets} = \textbf{\textit{find\_offsets}}(\textit{graph, inputWl, edgesPerThread, numThreads}) \label{getoffset-line-2} \\
}

\BlankLine
\Fn(){\textbf{sssp\_kernel}(graph, dist, inputWl, outputWl, offsets)}{

\textit{source = offsets[myid].NodeOffset} \\
\textit{ecurrent = offsets[myid].EdgeOffset} \\
\For{(\textit{edge = 0; edge $<$ edgesPerThread; ++edge})}{
  \While{ecurrent does not belong to source} {	\label{while-line} \tcc{check next node's edge}
	++\textit{offsets[myid].NodeOffset} \\
     	\textit{source = offsets[myid].NodeOffset} \\
     	\textit{offsets[myid].EdgeOffset = 0} \\
	\textit{ecurrent = offsets[myid].EdgeOffset + first edge index of source} \\
  }
%
   \textit{altdist = dist[ecurrent.source] + ecurrent.weight} \\
   \If{dist[ecurrent.destination] $>$ altdist}{
	\textit{dist[ecurrent.destination] = altdist} \\
	\textit{outputWl}.push(nodes with updated distance values) \\
   }
   \textit{offsets[myid].EdgeOffset++} \\
   \textit{ecurrent++} \\
}
}
\caption{SSSP Pseudocode illustrating Workload Decomposition}
\label{WD-cpu-pseudocode}
\end{small}
\end{algorithm}

\REM {
\begin{algorithm}
\begin{small}
\SetKwProg{Fn}{}{ \{}{\}}

\Fn(){CUDAKERNEL($dist, graph, inputWl, outputWl$)}{

Intialize $tid, start, end$ \;
$size$ = $numThreads$ = size of inputWl \;
$edgePerThr = ceil(prefixsum[size-1] / numThreads)$ \;
$source = $ node number located at $offsets[tid].NodeOffset$\;
$currentindex =$ startindex of $source$ in edge array of CSR $+ offsets[tid].Edgeoffset$ \;
$lastIndex =$ startIndex of $source$ in edge array of CSR $+$ outdegree of $source$ \;

\For{($edge=0; edge<edgePerThr; edge++$)}{
  \While{($currentindex >= lastIndex$)}{

    \tcc{if $currentIndex$ points to an edge of the next node}
     $offsets[tid].NodeOffset++$ \;
     $source = $ node number located at $offsets[tid].NodeOffset$\;
     $offsets[tid].EdgeOffset = 0$ \;
     $currentindex =$ startindex of $source$ in edge array of CSR $+ offsets[tid].Edgeoffset$ \;
   }
   Get $destination$ at $currentindex$ \;
   Initialize $weight$; $altDist = dist[source] + weight$ \;
   \If{$dist[destination] > altDist$}{
      Atomically update $dist[destination]$ with $altDist$ \;
      Atomically add node numbers of the updated nodes to $outputWl.Nodenumbers$ \;
      Add the outdegree of node to $outputWl.outgoing$ \;                   
   }
   $offsets[tid].EdgeOffset++$ \;
   $currentindex++$ \;
}
}
\caption{GPU Pseudocode for Workload Decomposition}
\label{WD-gpu-pseudocode}
\end{small}
\end{algorithm}
}

An advantage of workload decomposition is that it works with the CSR format and therefore, has a lower space complexity.
Assuming a conservative estimate of the number of nodes equal to half the number of edges, graphs of at least 350 million edges can be accommodated with the CSR format on the GPU.  
Also, since it distributes edges of a node across threads, it has a better load-balancing compared to a node-based distribution.
A drawback of the workload decomposition is that it can lead to uncoalesced accesses since a node's edges may get separated.
The method also incurs some overhead for the prefix sum operations, extra kernel-calls to obtain node offsets, and due to the atomic instructions required as a node may be operated upon by multiple threads.
Despite saving space compared to a COO format, workload decomposition requires extra space to store the node and edge offsets for each thread.

In our experiments, we observe that the limitations of workload decomposition affect its performance for large-diameter graphs (such as the road networks) 
but the method performs very well for scale-free graphs such as the social networks (Section~\ref{exp_res}).

\subsection{Node Splitting}\label{node-splitting}
\begin{figure}
\centering
\includegraphics[scale=0.48]{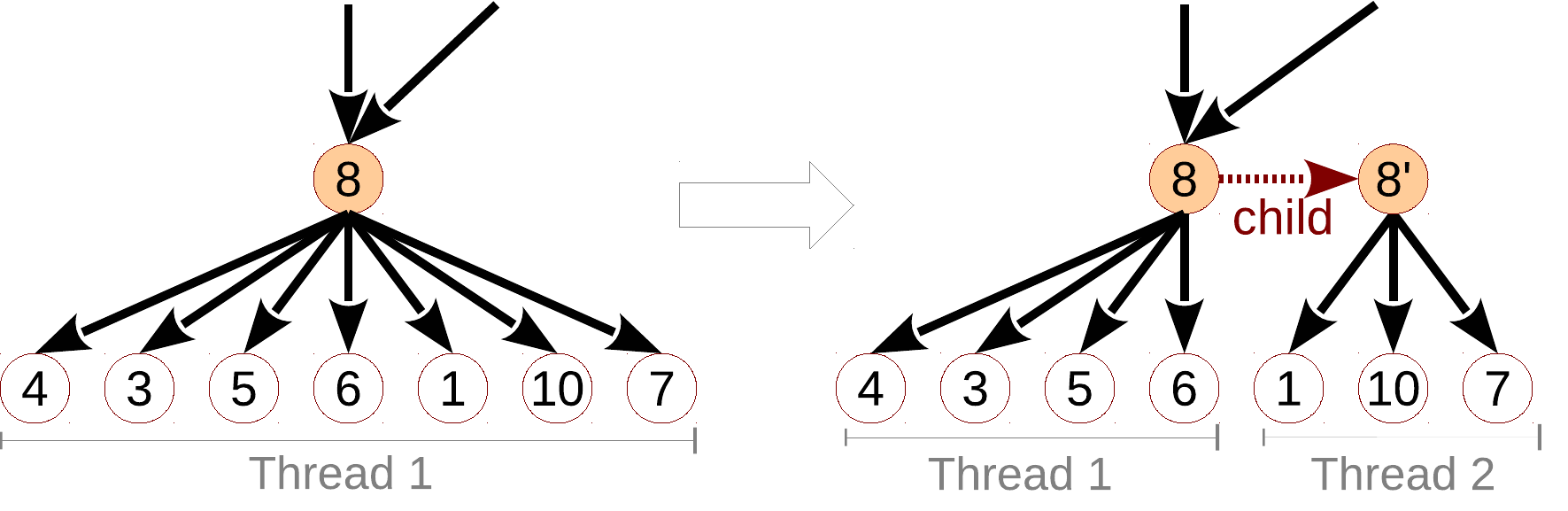}
\caption{Node Splitting (maximum outdegree threshold MDT = 4)}
\label{fig:ns}
\end{figure}

The second approach we propose is based on changing the graph structure itself to balance the load.
Since the root cause of the load-imbalance is skewed outdegree distribution across graph nodes, node splitting preprocesses the graph to split each high-degree node into multiple low-degree child-nodes.
This approach is implemented as follows.
We define an input parameter called maximum-out-degree-threshold (MDT). 
If a node's outdegree is more than MDT, then the node is split into $\lceil\frac{outdegree}{MDT}\rceil$ nodes, with the outgoing edges of the node distributed evenly among the original (parent) and the split (child) nodes. 
For example, Figure~\ref{fig:ns} depicts our node splitting approach, where a node 8 is split into two nodes, 8 (parent) and 8' (child) which share the outgoing edges.
Multiple child nodes can get formed from a parent node. 
Note that the graph now does not contain the original high-degree node.
By repeating this splitting procedure for each of the high-degree graph nodes, we can obtain another graph with maximum degree bounded by MDT.

As evident, node-splitting approach has the advantage that it can work with the space-efficient CSR representation.
Since node-splitting creates duplicates of a node, incoming edges to the original node need special consideration.
To address this issue without increasing any existing node degrees, we maintain the incoming edges of a node only for the parent node.
The parent node, in turn, keeps track of its children.
The algorithm is modified to reflect the attributes (distance in case SSSP) of a parent node onto its children.
This strategy ensures that no new edges get added to the graph (parent-child relationship does not interfere with the normal graph edges).

\begin{figure*}
\centering
\includegraphics[scale=0.4]{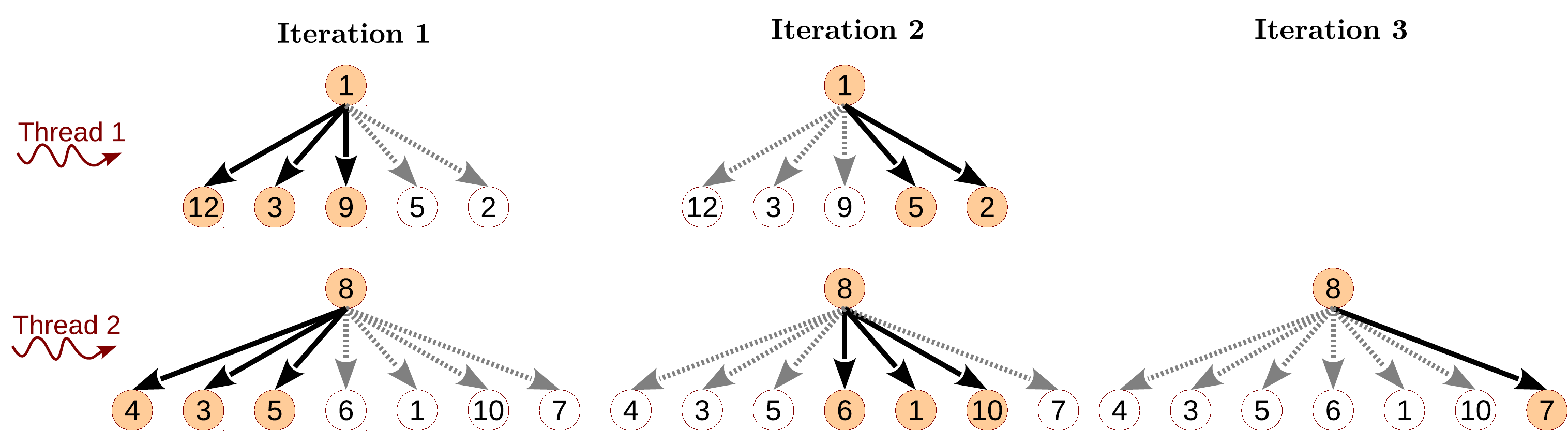}
\caption{Hierarchical Processing (MDT = 3, i.e., three edges per node are processed in each iteration)}
\label{hp-illustration}
\end{figure*}

\vspace{0.1in}
\noindent
{\bf Automatic Determination of Node Splitting Threshold:}
A salient feature of our node splitting strategy is to automatically determine the threshold MDT for node splitting.
Obvious methods based on a threshold or max-degree etc. do not work in general.
For instance, we cannot fix the value of MDT to a constant, since it is unsuitable across graphs and degree distributions.
We cannot also fix MDT to the maximum degree in the graph, since there could be a big skew in the degree distribution with a few very high degree nodes and a large number of medium degree nodes.
A better alternative is to use the difference between the average degree and the maximum degree in the graph; but such a function would be influenced by the graph size.
Another constraint is that the number of nodes to be split should be minimum possible, to reduce the splitting (and processing) cost.
To account for these issues, 
we use a histogram based method in which we use \textit{HistogramBinCount} number of bins representing the ranges of outdegrees of the nodes in the original graph. 
The number of bins is given as an input parameter to our algorithm. 
We then find the distributions of the outdegrees across the bins. 
We find the bin or range with the maximum height, i.e., the range of outdegrees for which the graph has the maximum number of nodes. 
Let \textit{binIndex} be the index of this bin. 
We then find the maximum degree threshold \textit{MDT} for the outdegrees as \textit{(binIndex/HistogramBinCount)$\times$ maxDegree}, where \textit{maxDegree} is the maximum outdegree in the graph. 
Our node-splitting algorithm then finds the nodes in the graph with outdegrees greater than \textit{MDT} and splits them into child nodes such that each parent and the child nodes will have a maximum of \textit{MDT} outdegrees. 
Our histogram approach of finding \textit{MDT} attempts to maximize the number of nodes (parent and child) with \textit{MDT} outdegrees. 
By choosing the bin with the maximum height in which the nodes already have their outdegrees closer to \textit{MDT}, our algorithm minimizes the amount of splitting. 
Since the histogram-based method considers degree distributions of a graph to achieve load balancing, it can be applied to different graphs with different kinds of distributions including highly skewed distributions.

An advantage of the node splitting approach is that it continues to work with the space-efficient CSR format.
Another advantage over workload-decomposition (Section~\ref{workload-decomposition}) is that all the edges of a node are processed by the same thread, reducing bookkeeping and improving the scope for memory coalescing.
The primary disadvantage of node splitting is that it results in extra atomic operations to update the child nodes whenever the parent node gets updated.
A secondary disadvantage is the overhead of computing the histogram to find the MDT.
One may wonder that the strategy has a space overhead due to explicit splitting, but we found in our experiments that less than 5\% of the nodes undergo split, resulting in negligible space overhead.

As we observe in our experiments (Section~\ref{exp_res}), node-splitting provides considerably better load-balancing.
In addition, it provides comparable performance for large diameter graphs (such as road networks); but it has a high overhead for power-law degree distribution graphs.

\subsection{Hierarchical Processing}\label{hierarchical-processing}
Hiearchical processing performs a time-decomposition of the workload.
It achieves this by partitioning the main (super) worklist into several sub-worklists. 
If the sub-worklist is large, it can be further partitioned into sub-sub-worklists, and so on. 
This builds a hierarchy of worklists. 
The depth of this hierarchy is tunable, and we utilize the histogram-based approach in node-splitting (Section~\ref{node-splitting}) for finding the maximum degree threshold (MDT) which determines when to split a worklist into sub-worklists.

An iteration for processing a node worklist is composed of sub-iterations. 
In each sub-iteration, a sublist consisting of nodes remaining to be processed from the super-worklist is formed, and a GPU kernel is invoked to process this sublist. 
Each GPU thread considers a set of nodes in the sublist, processing only up to MDT unprocessed outgoing edges of these nodes. 
Thus, all the threads corresponding to the kernel invocation of the sub-iteration are load-balanced within this threshold. 
The nodes in the sublist with the number of unprocessed outgoing edges less than or equal to MDT will be processed. 
The next sublist with a reduced set of nodes will be processed in the next sub-iteration. 
This continues until all the nodes in the super list are processed before processing the next super list in the next iteration. 

Figure~\ref{hp-illustration} illustrates the hierarchical processing of sublists within an iteration for an input graph. 
The super worklist in the figure contains two nodes 1 and 8 with five and seven outgoing edges respectively. 
In each kernel for a sub-iteration, two threads are spawned for processing the two nodes.
Let MDT be $3$. Then each thread in a sub-iteration processes maximum three edges.


The sizes of the sublists decrease over the sub-iterations due to the removal of the processed nodes. 
Since the GPU kernel is spawned using a node-parallel approach in which the number of GPU threads is proportional to the size of the sublist, the reduction of the sublist size can result in a situation where the GPU kernel is invoked with a small number of threads to process a small number of nodes in the sublist. 
This will result in very low GPU occupancy. 
For example, consider a situation where, if after a few sub iterations, only one node remains to be processed and this node has $100$ edges remaining to be relaxed. 
If the MDT is $5$, twenty more sub-iterations will invoke twenty more GPU kernels successively, each spawning one GPU thread to process $5$ edges of the node. 
To avoid this situation, our strategy switches to the workload-decomposition technique when the sublist size becomes smaller than a threshold.
A natural threshold is governed by the GPU kernel block size which, in our experiments, is set to $1024$. 
We also switch to workload decomposition for processing the super worklist at the beginning of the top-level iterations when the size of the super list becomes smaller than the block size.

The fundamental advantage of the hierarchical processing strategy over node-splitting is that it avoids the space and time complexity needed for creation of new nodes. 
By following a hybrid method of using workload-decomposition for small number of nodes and using the technique of sub-iterations for larger number of nodes, it combines the advantages of multiple approaches. 
However, the hierarchical processing method incurs extra overhead due to additional kernel invocations corresponding to the sub-iterations. 
The method also incurs increased space complexity and atomic operations for the sub worklists.

In our experiments (Section~\ref{exp_res}) we found that despite its overheads, hierarchical processing is a scalable mechanism.
For larger graphs in our experimental suite where other proposed strategies fail to execute due to insufficient memory requirment, hierarchical processing
successfully completes execution offering good benefits in terms of load-balancing.

Table \ref{strategies-summary} summarizes the advantages and disadvantages of the different load balancing strategies.

\begin{table*}
 \centering
 \footnotesize
 \caption{Advantages and Disadvantages of the Load Balancing Strategies}
 \begin{tabular}{|c|p{1.45in}|p{2.45in}|p{2.55in}|}
  \hline\hline
  & {\em Strategy} & {\em Advantages} & {\em Disadvantages} \\
  \hline\hline
  \parbox[t]{2mm}{\multirow{2}{*}{\rotatebox{90}{Existing}}} &
  Node-based Distribution (BS) &
   \parbox{2.5in} {
	\begin{itemize}
	\item Simple to implement (static)
	\item  Works with CSR graph format
   	\end{itemize} 
   } &
   \parbox{2.5in} {
   	\begin{itemize}
	\item High load-imbalance
   	\end{itemize} 
   } \\\cline{2-4}
   &
  Edge-based Distribution (EP) & 
   \parbox{2.5in} {
   	\begin{itemize}
	\item Implicit load balancing 
	\item Simple to implement (static)
   	\end{itemize} 
   } &
  \parbox{2.5in}{
  \begin{itemize}
   \item Large space complexity for COO representation
   \item Explosion in worklist size, worklist condensing overhead, large memory consumption
   \item Requires the kernel operation to be distributive
  \end{itemize}
  } \\ \hline
  \parbox[t]{2mm}{\multirow{3}{*}{\rotatebox{90}{Proposed}}} &
  Workload Decomposition (WD) & 
  \parbox{2.5in}{
  \begin{itemize}
   \item Larger graphs can be processed
   \item Space decomposition is easy to implement
  \end{itemize}
  } & 
  \parbox{2.5in}{
  \begin{itemize}
   \item Atomic operations for updating same out-going edges by multiple threads
   \item Overheads for prefix sum and offset computations, extra space for offsets
   \item  Uncoalesced data access on the GPU
  \end{itemize}
  } \\ \cline{2-4}
  &
  Node Splitting (NS) & 
  \parbox{2.5in}{
   \begin{itemize}
	\item Graph algorithm does not require modification
   \end{itemize} 
  } &
  \parbox{2.5in}{
  \begin{itemize} 
   \item Additional space and time complexity for new nodes
   \item Extra atomic operations for updating child nodes
   \item  Overhead for MDT finding
  \end{itemize}
  } \\ \cline{2-4}
  &
  Hierarchical Processing (HP) & 
  \parbox{2.5in}{
  \begin{itemize}
   \item Performs well for large graphs
   \item  A thread processes only one node without forming child nodes
   \item  Hybrid method for switching to workload decomposition strategy for small super and sub worklists
  \end{itemize}
  } & 
  \parbox{2.5in}{
  \begin{itemize}
   \item Sub lists result in additional space and atomics
   \item Multiple kernel calls
  \end{itemize}
  } \\ \hline\hline
 \end{tabular}
 \label{strategies-summary}
\end{table*}

\section{Experiments and Results}
\label{exp_res}

To assess the effectiveness of our proposed techniques, we embed them in the implementation of two graph algorithms: breadth-first search computation (BFS) and single-source shortest paths computation (SSSP).
Both these algorithms are fundamental to several domains and form the building block for several interesting applications.
We compare our implementations against the LonestarGPU benchmark implementations~\cite{lonestargpu-web},
which use a node-based task-distribution.
We used LonestarGPU-1.02 version, an older version available at the time of our work. 
For our experiments, we use both synthetically generated graphs as well as the real-world graphs.
The synthetic ones are the RMAT graphs based on the recursive matrix model~\cite{gtgraph} and random graphs based on the Erd\H{o}s-R\'enyi model (denoted as ER*). 
Both are generated using GTgraph~\cite{gtgraph}.
For real-world graphs, we use the USA road networks (for West, Florida and overall USA).
To assess scalability, we include three relatively larger graphs obtained using the graph generation tool available in the Graph500 benchmark~\cite{graph500-web}. 
The tool accepts three parameters: number of nodes, number of edges and a seed value for random number generation, and generates a corresponding graph.
Depending upon the seed value, the graph connectivity differs.
The properties of all the graphs are given in Table~\ref{graph_properties}. 
The last column of the table represents the amount of load imbalance in the form of the skewness in the outdegree distribution of the nodes.
This is indicated in terms of the maximum, average and standard deviations ($\sigma$) of their outdegrees.

We observe in Table~\ref{graph_properties} that Graph500 and RMAT graphs have a high maximum degree as well as a lot of variance in the number of outdegrees.
RMAT graphs are also characterized by \textit{small-world property} due to their low diameter.
In contrast, the road networks have very small maximum degree and little variation in the outdegree distribution. 
They have large diameters (not shown) in comparison to the RMAT and ER graphs.
ER graphs have a random distribution of edges in the graph and have a higher max-degree as well as the standard deviation than road networks.
However, they do not have large diameters as in the case of road networks, nor do they exhibit the small-world property.
It should be noted that despite this variance, the average degrees of all the graphs remain comparable.
Together these graphs test various aspects of our strategies.

\begin{table}
\centering
\caption{Graphs Used in our Experiments}
\footnotesize
\begin{tabular}{|r|r|r|rrr|}
\hline
{\em Graph} 	& {\em Nodes} 		& {\em Edges} 		& \multicolumn{3}{c|}{\em Outdegrees}	\\
 		& {\em (Million)}	& {\em (Million)}	& {\em Max}	& {\em Avg}	& $\sigma$ \\
\hline\hline
rmat20 & 1.05 & 8.26 & 1,181 & 8 & 177.40 \\
\hline
road-FLA & 1.07 & 2.71 & 8 & 3 & 2.45 \\
road-W & 6.26 & 15.12 & 9 & 4 & 2.74 \\
road-USA & 23.95 & 57.71 & 9 & 3 & 2.74 \\
\hline
ER20 & 1.05 & 4.19 & 15 & 4 & 4.47 \\
ER23 & 8.39 & 33.55 & 10 & 3 & 4.46 \\
\hline
Graph500 & 16.78 & 335.00 & 924,000 & 20 & 20,900 \\
(three graphs) &  &  &  &  &  \\
\hline
\end{tabular}
\label{graph_properties}
\end{table}

We implemented all our proposed strategies using CUDA.
Our experiments were performed on a Kepler-based GPU system. 
The GPU has a Tesla K20c architecture with 13 SMXs each having 192 CUDA cores (2,496 CUDA cores totally) with 4.66 GB of main memory, 1 MB of L2 cache and 64 KB of registers per SM.  
It has a configurable 64 KB of fast memory per SMX that is split between the L1 data cache and shared memory. 
The programs have been compiled using \textit{nvcc} version 5.0 with \textit{-O3 -arch=sm\_35} flags. 
The CPU is a hex-core Intel Xeon E5-2620 2.0 GHz workstation with CentOS 6.4, 16 GB RAM and 500 GB hard disk.

\newcommand{\figurewidth}{0.25}

\begin {figure*}
\centering
\subfigure[Low Diameter Graphs]{
  \includegraphics[width=\figurewidth\linewidth]{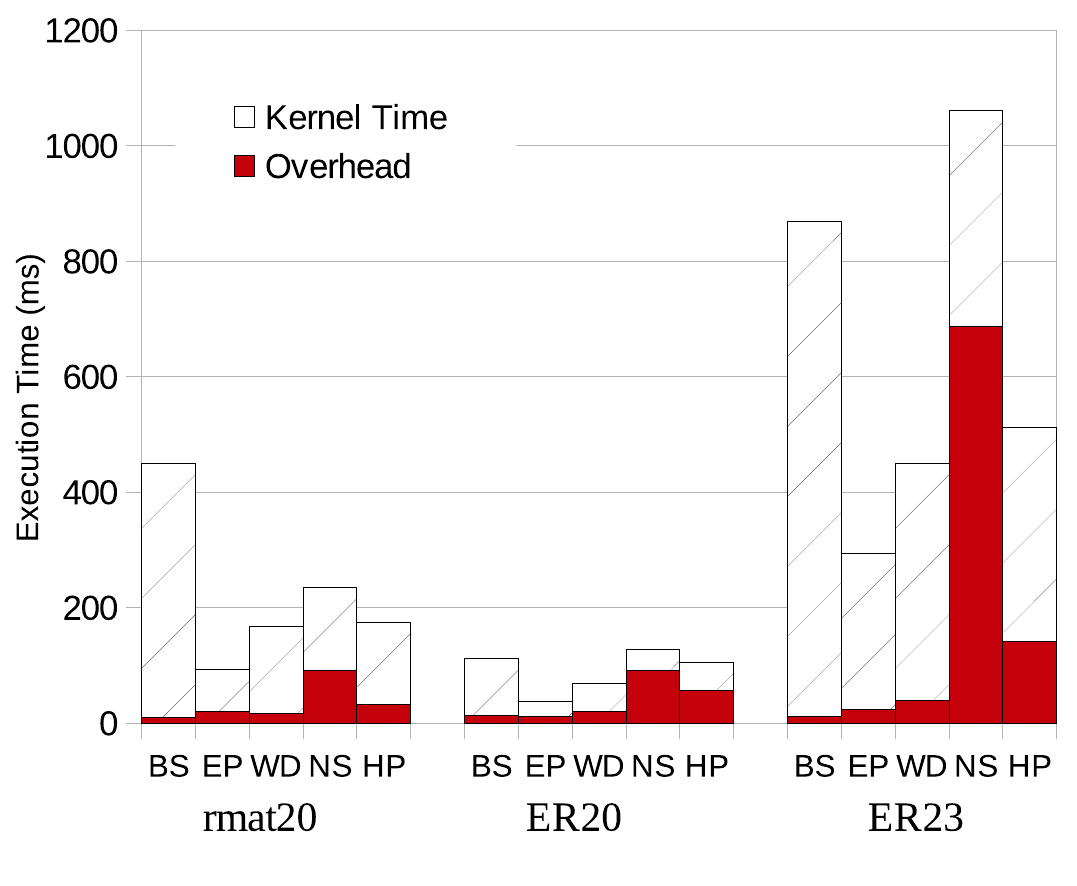}
  \label{rgraphs_sssp}
}
\subfigure[Road Networks]{
  \includegraphics[width=\figurewidth\linewidth]{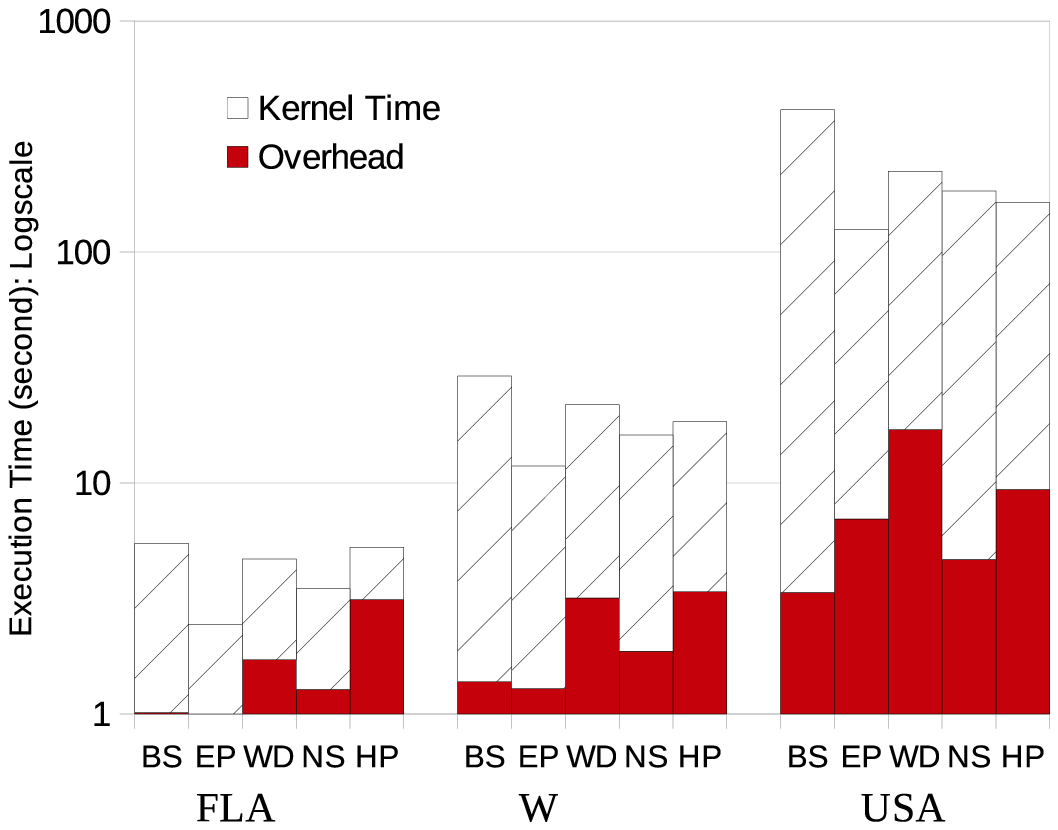}
  \label{USA-road-d_FLA_sssp}
  \label{USA-road-d_W_sssp}
  \label{USA-road-d_USA_sssp}
}
\subfigure[Graph500 Graphs]{
  \includegraphics[width=\figurewidth\linewidth]{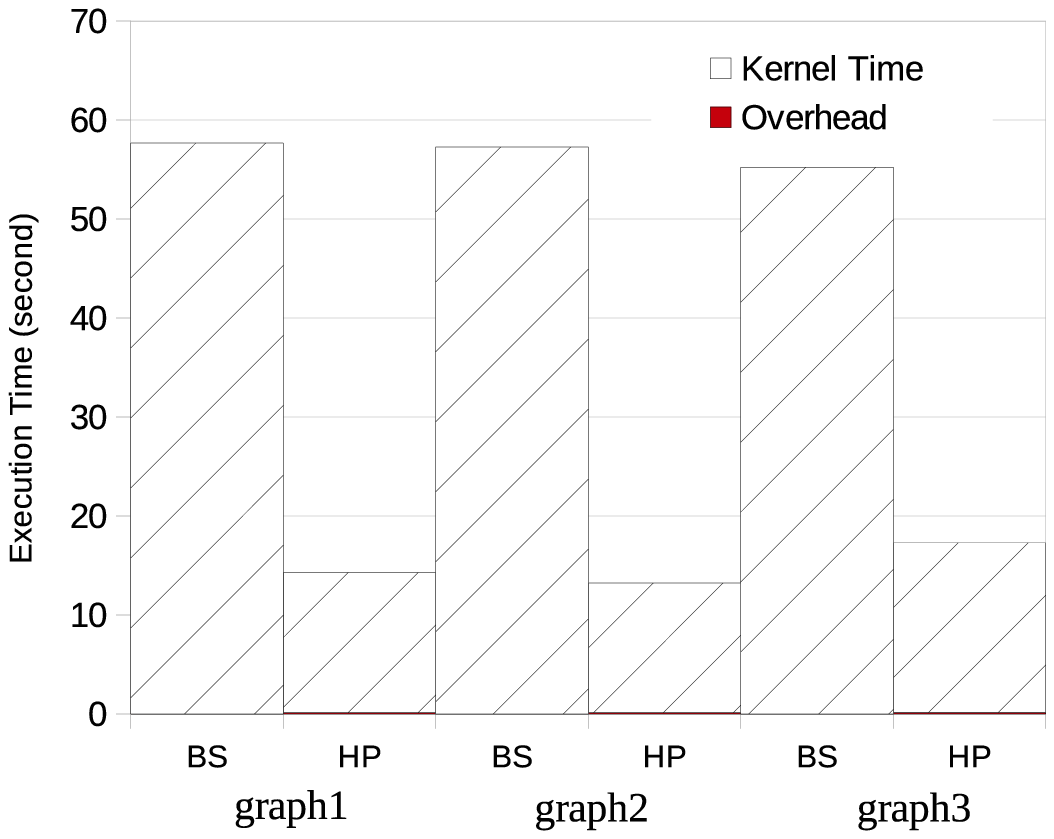}
  \label{very_large_sssp}
}
\caption{Comparison of Load Balancing Strategies for \textbf{SSSP}} 
\label{overall-comparisons-sssp}
\end{figure*}

\begin {figure*}
\centering
\subfigure[Low Diameter Graphs]{
  \includegraphics[width=\figurewidth\linewidth]{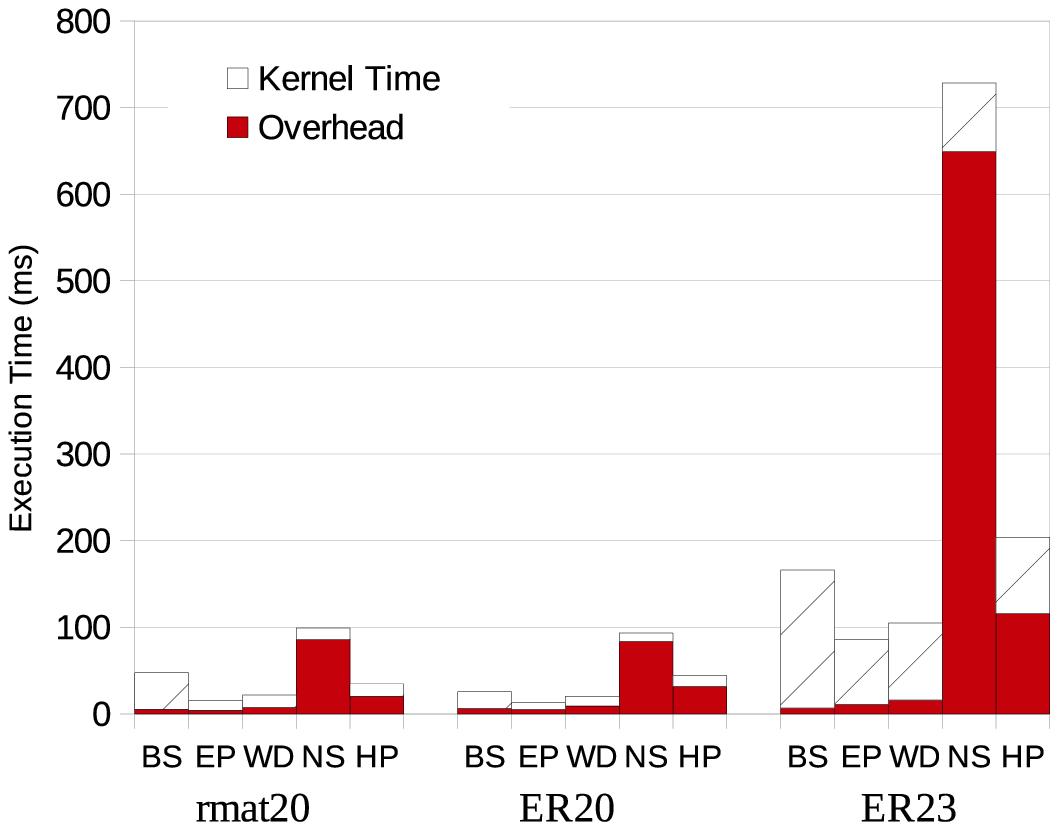}
  \label{rgraphs_bfs}
}
\subfigure[Road Networks]{
  \includegraphics[width=\figurewidth\linewidth]{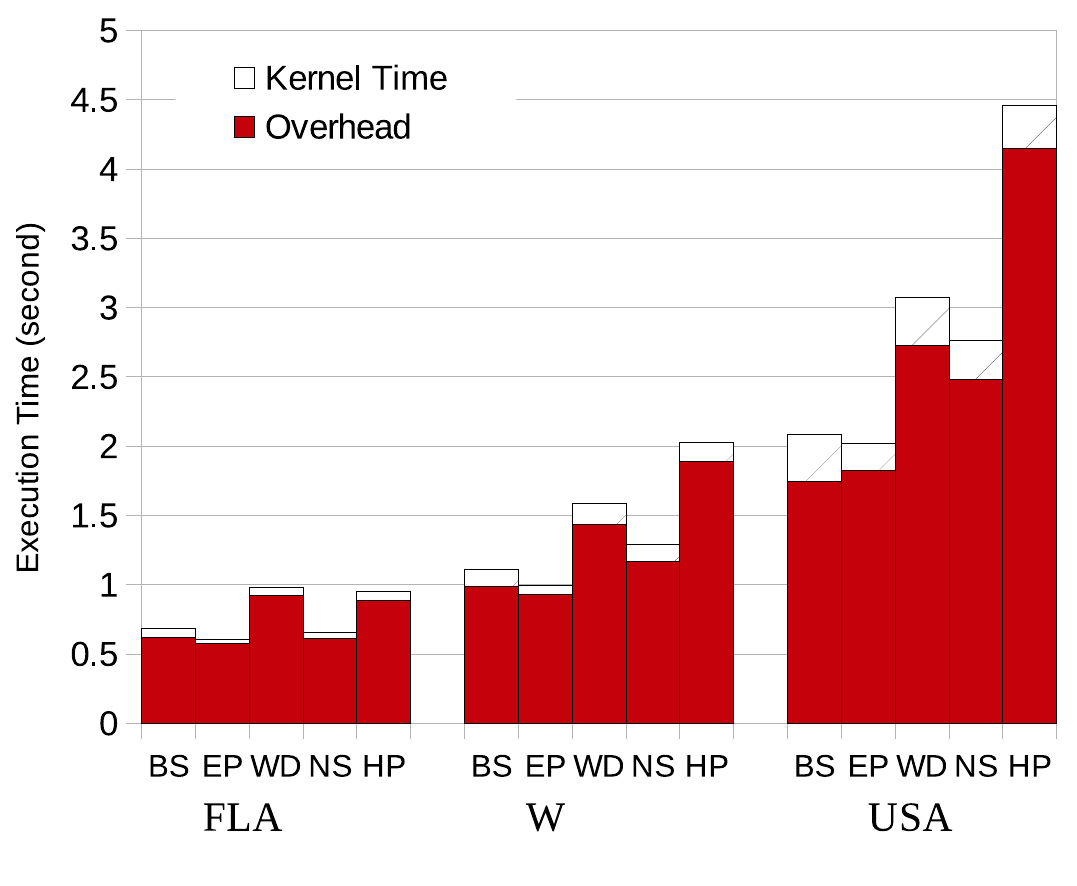}
  \label{USA-road-d_FLA_bfs}
  \label{USA-road-d_W_bfs}
  \label{USA-road-d_USA_bfs}
}
\subfigure[Graph500 Graphs]{
  \includegraphics[width=\figurewidth\linewidth]{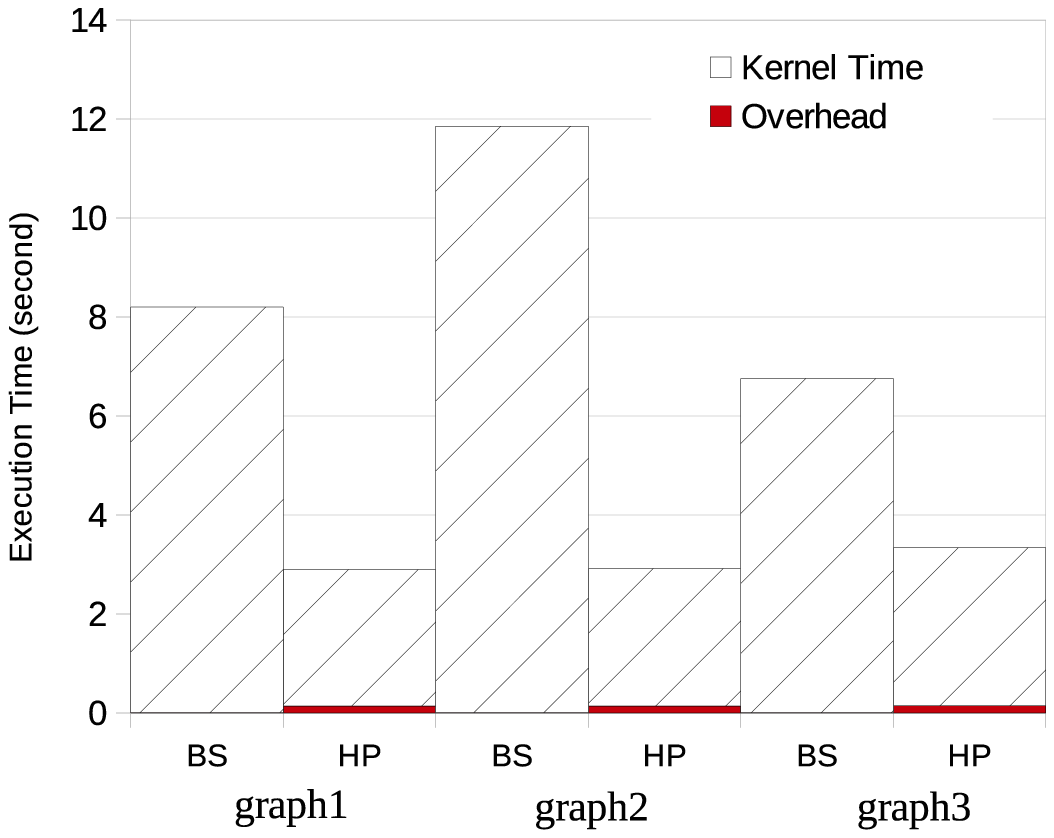}
  \label{very_large_bfs}
}
\caption{Comparison of Load Balancing Strategies for \textbf{BFS}}
\label{overall-comparisons-bfs}
\label{very_large_graph_results}
\end{figure*}

\REM {
\begin {figure}
\centering
\subfigure[Synthetic Graphs]{
  \includegraphics[scale=0.2]{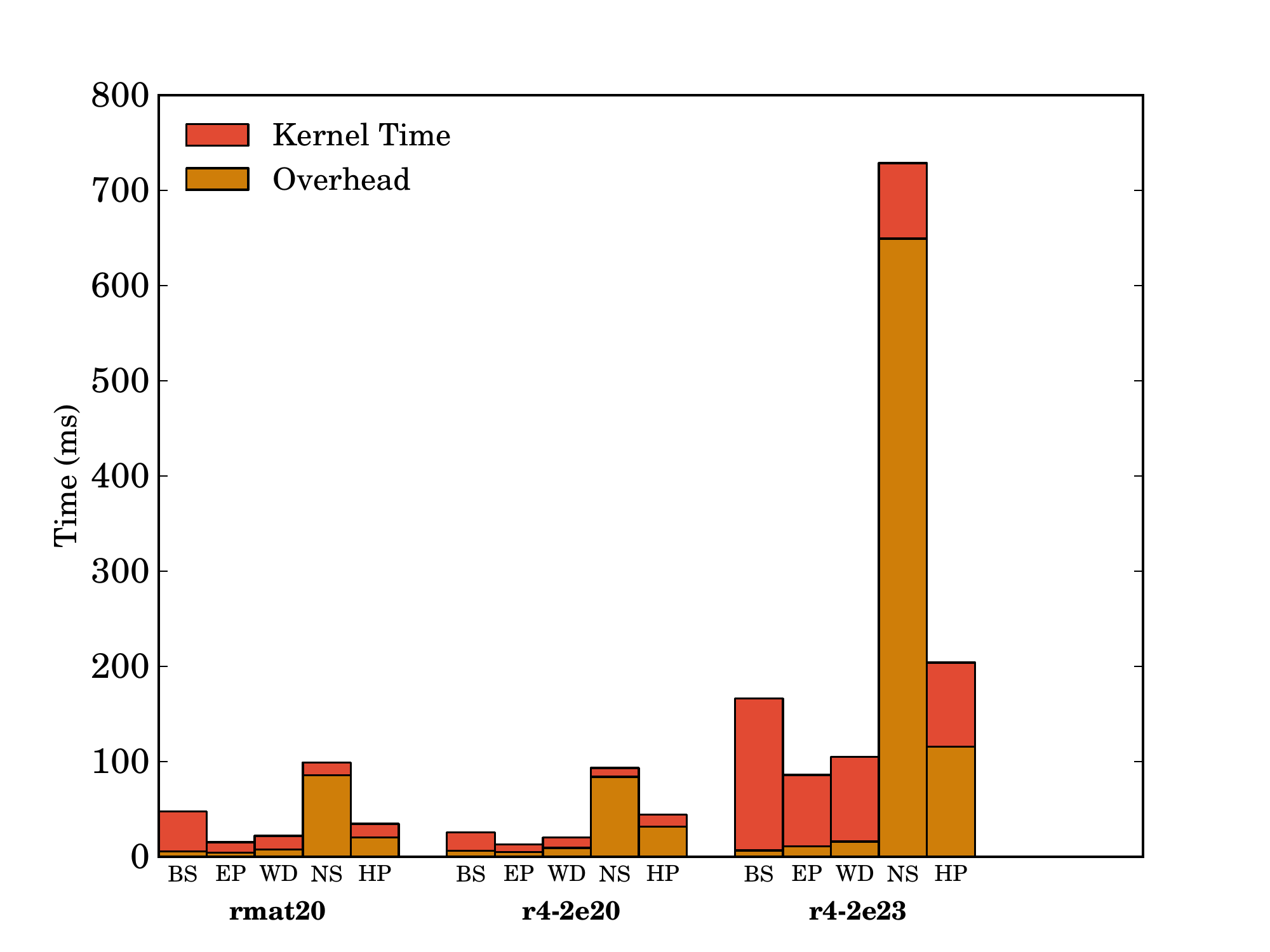}
  \label{rgraphs_bfs}
}
\subfigure[road-FLA]{
  \includegraphics[scale=0.2]{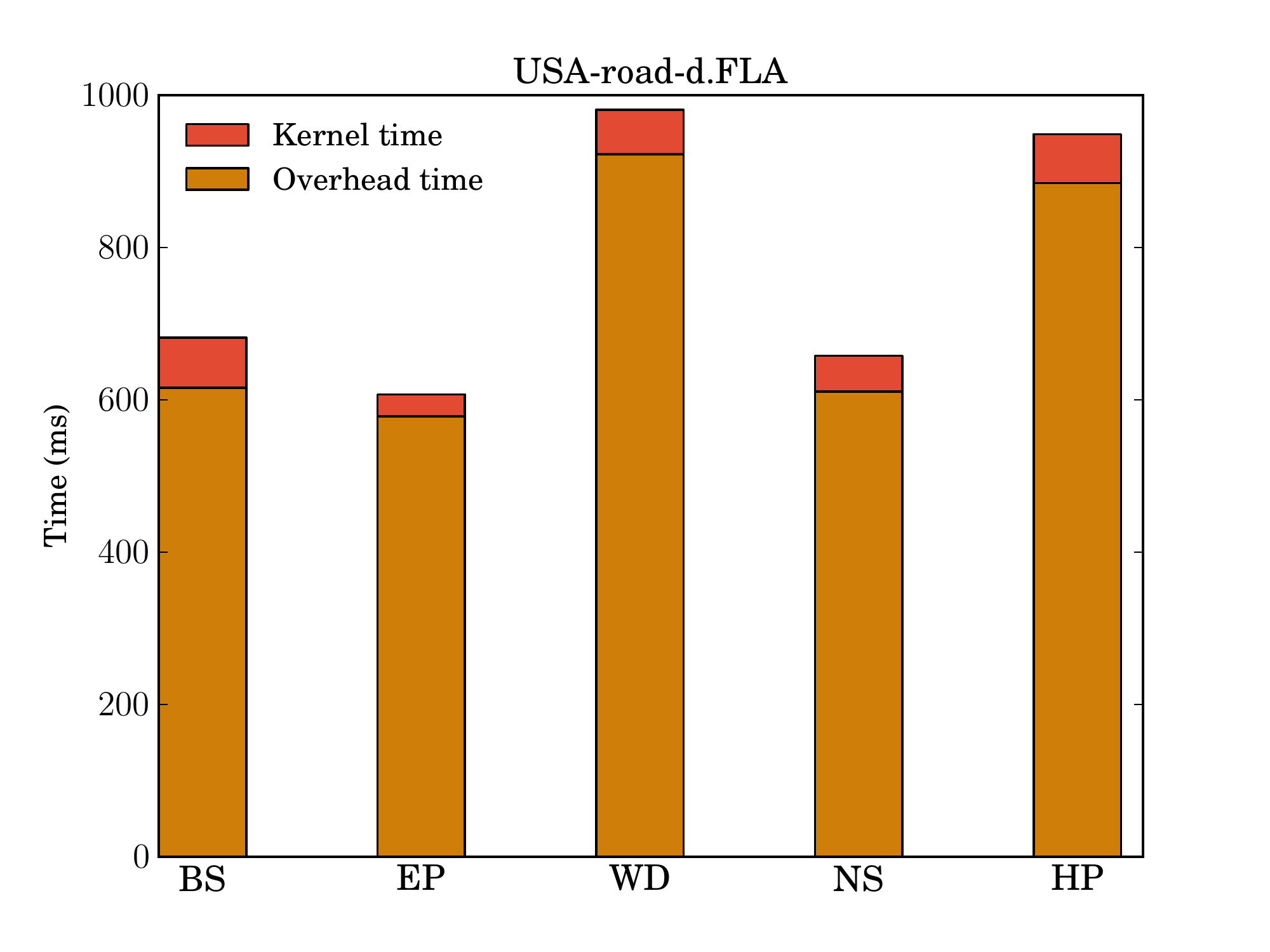}
  \label{USA-road-d_FLA_bfs}
}
\subfigure[road-W]{
  \includegraphics[scale=0.2]{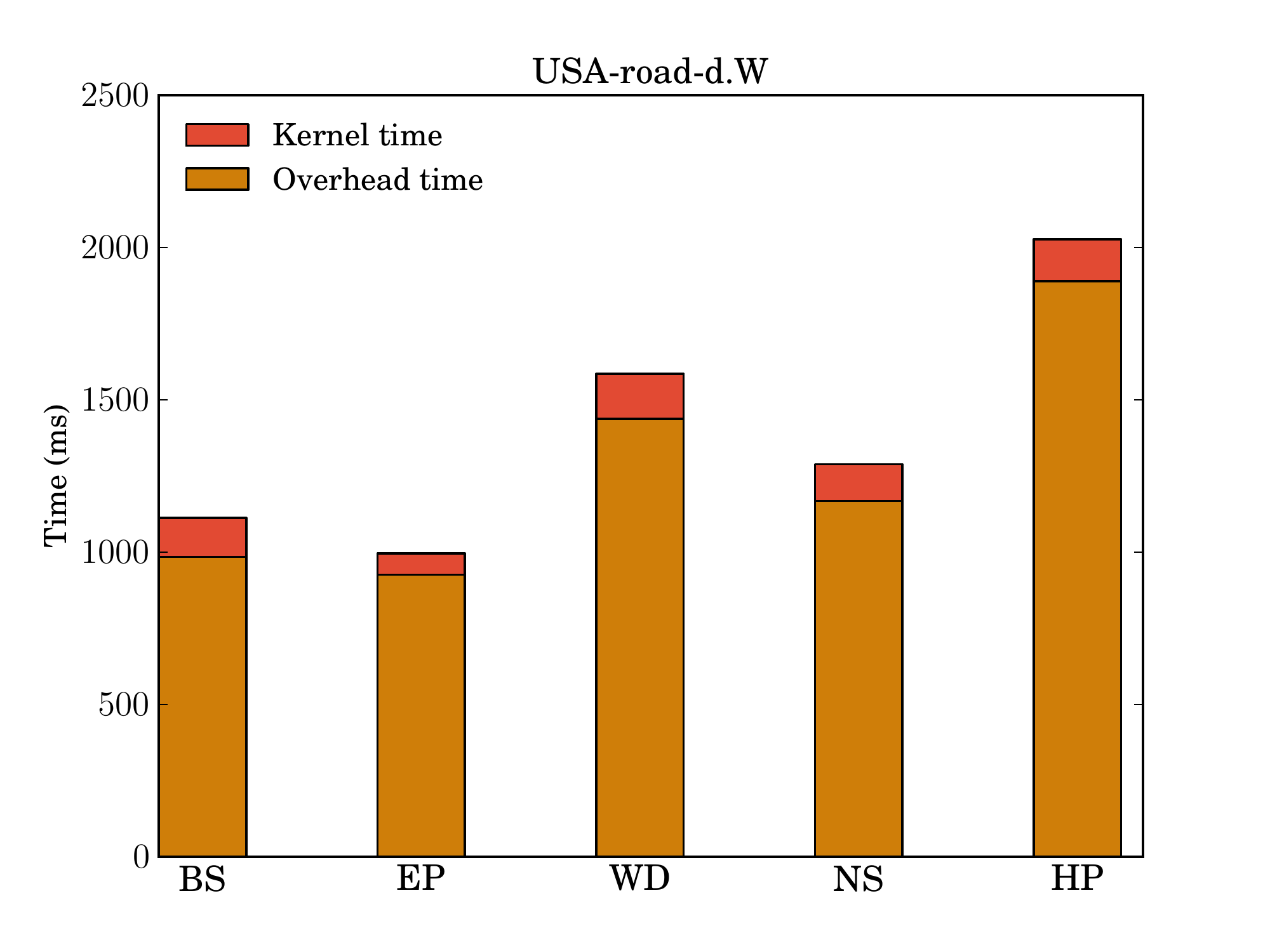}
  \label{USA-road-d_W_bfs}
}
\subfigure[road-USA]{
  \includegraphics[scale=0.2]{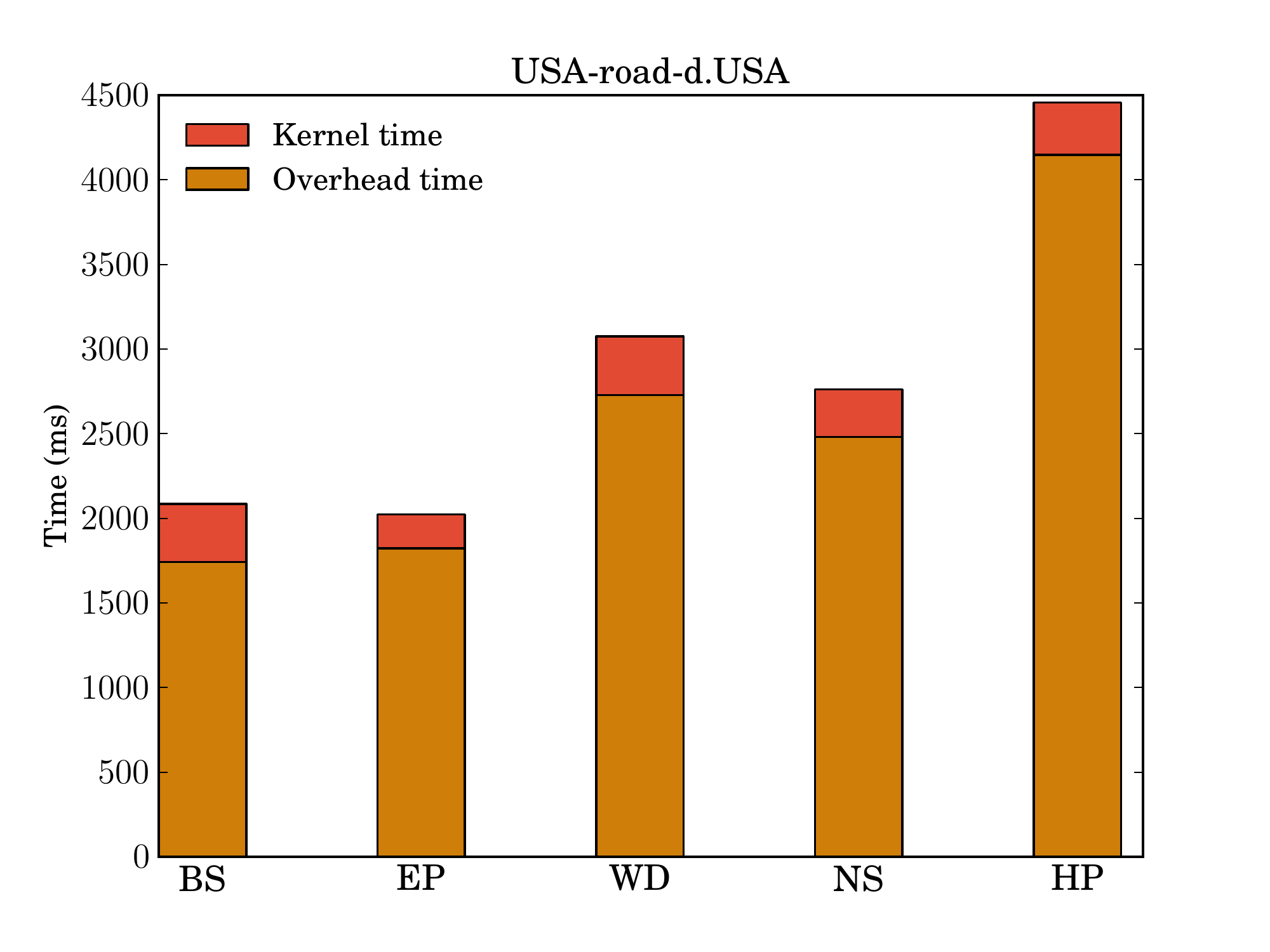}
  \label{USA-road-d_USA_bfs}
}
\caption{Comparison of Load Balancing Strategies - BFS}
\label{overall-comparisons-bfs}
\end{figure}
}

\subsection{Performance Comparison of Strategies}\label{expt:exectime}

In this section, we compare the various strategies in terms of performance or execution times. The strategies are denoted as BS for LoneStar GPU baseline version that implements node-based task-distribution, EP for edge-based distribution, WD for workload decomposition, NS for node-splitting, and HP for hierarchical processing. We split the overall execution time into useful kernel time and the overhead associated with implementing a strategy. The overheads encompass all the corresponding initializations, extra kernel invocations and bookkeeping. Note that BS also has an overhead component.

Figure \ref{overall-comparisons-sssp} shows the comparison results for SSSP. We find that all our strategies perform significantly better than the baseline (BS) method in almost all cases, for graphs with small as well as large diameters. This is because in SSSP, especially for the graphs with large diameters, the kernel times dominate the overheads (unlike in BFS, discussed below).  This shows that our load balancing strategies in particular, and load balancing in general, are fruitful for applications that perform even a reasonable amount of computations. We believe the techniques would be more useful for computation-intensive irregular applications. The edge-based parallelism (EP) method performs the best, giving 60--80\% smaller execution times than the baseline. Unfortunately, due to its high storage requirement, EP is unable to run on larger graphs such as Graph500. Among the node-based strategies, workload decomposition (WD) method performs the best for graphs with highly skewed or random degree distribution. For such graphs (RMAT and ER), the node splitting (NS) performs the worst since its node creation overhead coupled with highly skewed degree distribution dominates the kernel times. However, when the deviation in the size of the neighborhood is less, the NS method performs the best among the node-based strategies since its node creation overhead is a one-time cost and is amortized by relatively large total kernel execution times.

Hierarchical processing (HP) performs in between the WD and NS methods for smaller graphs. However, the main advantage of HP is seen in dealing with larger graphs such as Graph500. At the time of writing this paper, we were able to execute only the HP strategy of the three load balancing strategies (WD,NS and HP) for these large graphs.
As mentioned, the edge-based parallelism (EP) has a high storage complexity related to storing the edges and hence cannot be executed for these large graphs.
We find that the HP method gives large improvements resulting in 48-75\% reduction in execution times for these large graphs. Thus the HP method will have larger importance as we explore real-world BigData graphs.

\REM {
\begin {figure}
\centering
\subfigure[BFS]{
  \includegraphics[scale=0.2]{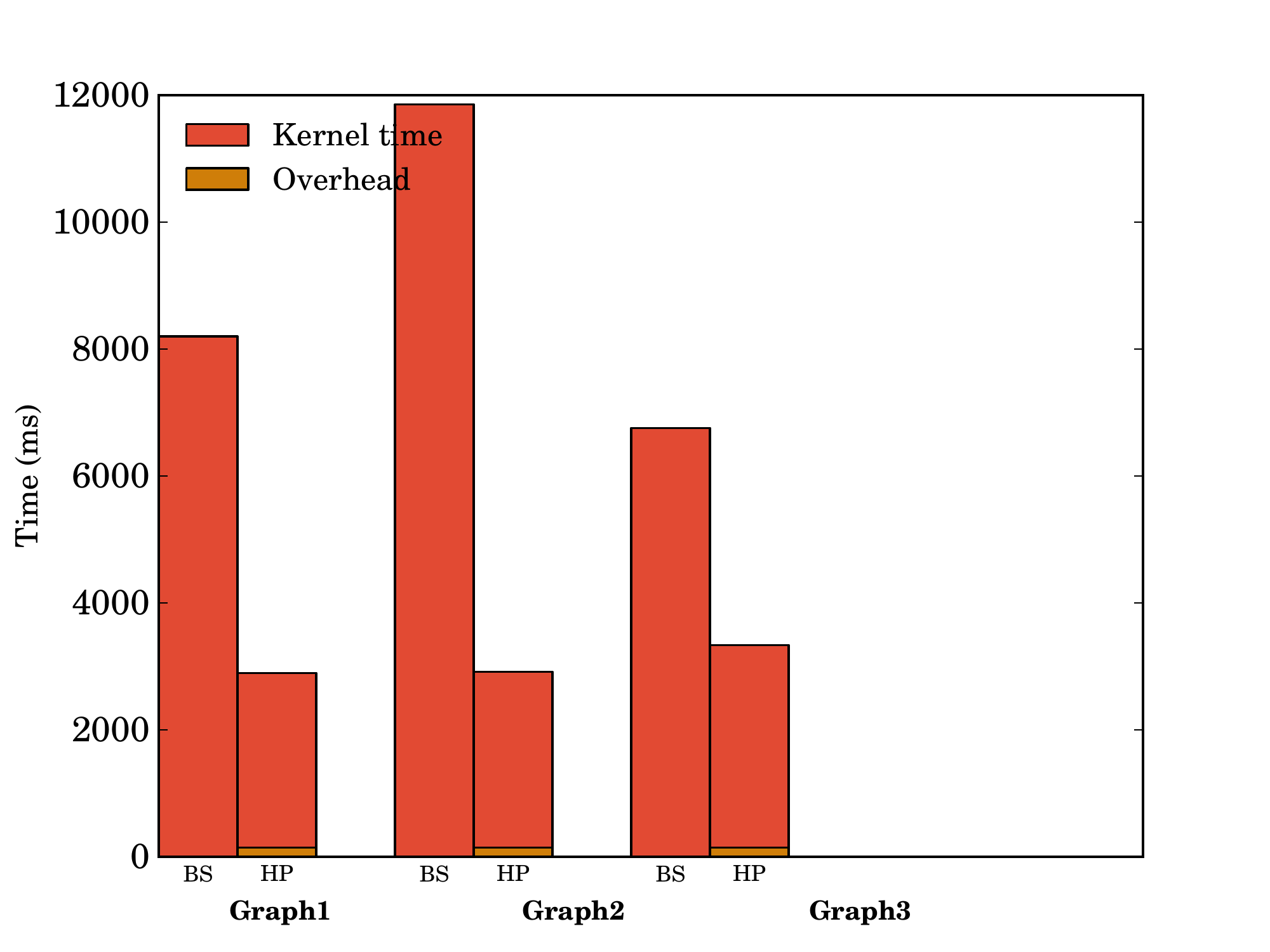}
  \label{very_large_bfs}
}
\subfigure[SSSP]{
  \includegraphics[scale=0.2]{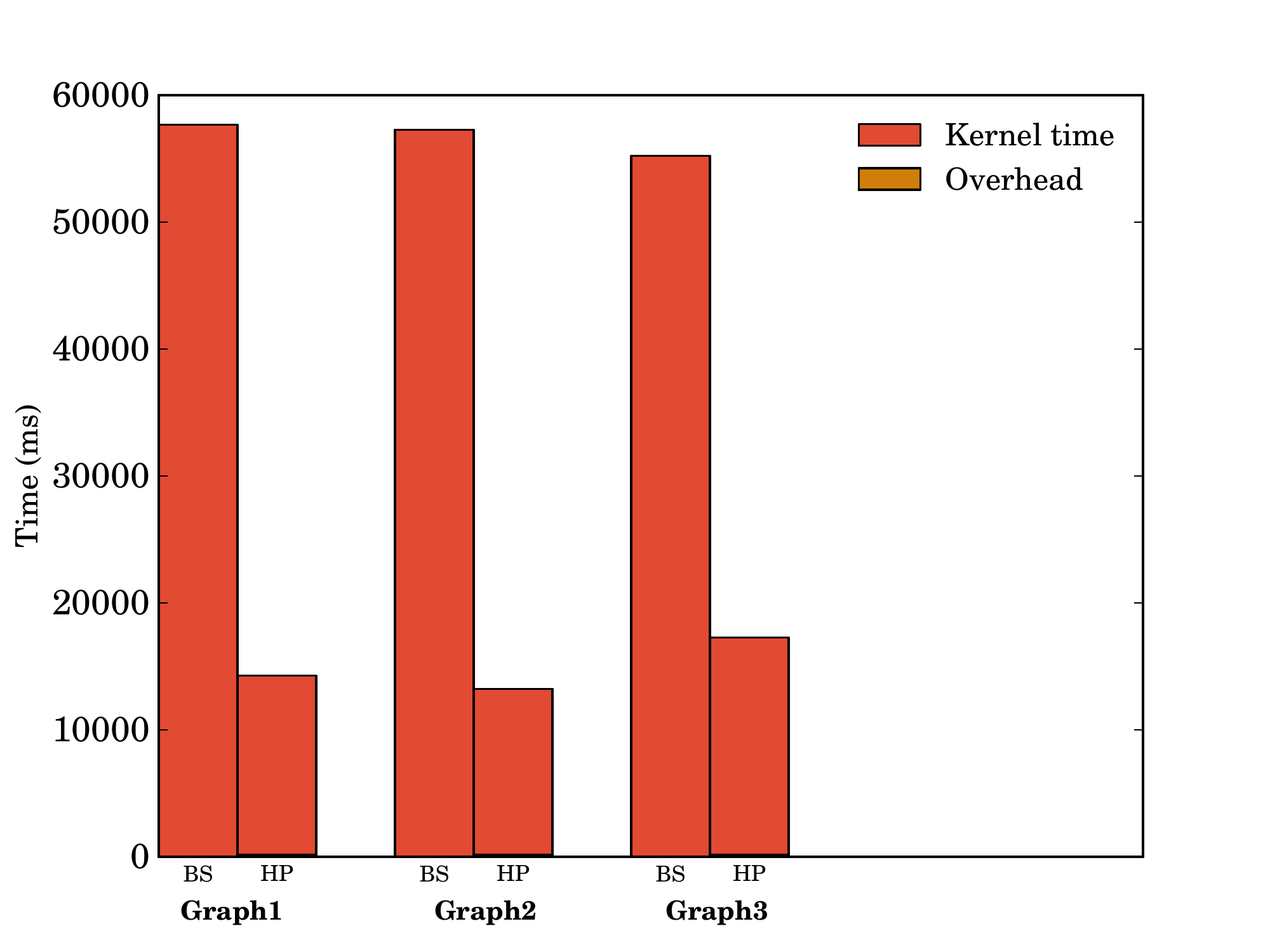}
  \label{very_large_sssp}
}
\caption{Results for Very Large Graphs}
\label{very_large_graph_results}
\end{figure}
}

\REM {
Following are the overheads reported for the different strategies: \\
1. Baseline (BS): initialization of worklist, and total time for the main loop excluding GPU kernel executions. \\
2. Edge-based parallelism (EP): initialization of worklist, and total time for the main loop excluding GPU kernel executions. \\
3. Workload Decomposition (WD): initialization of worklist, CUDA kernel for obtaining the offsets before the main loop, and total time for the main loop excluding GPU kernel executions. \\
4. Node Splitting (NS): finding node-splitting level using histogram approach and 10 bins, initialization of worklist, and total time for the main loop excluding GPU kernel executions. \\
5. Hierarchical processing (HP): finding node-splitting level using histogram approach and 10 bins, creating space for extra worklists, initialization of worklist, creating GPU memory for allocating worklists and total time for the main loop excluding GPU kernel executions.
}

Figure \ref{overall-comparisons-bfs} shows the comparison results for BFS. It is noteworthy that BFS is a memory-bound kernel, and it performs only a little computation. Therefore, we observe the associated overheads are large in general, unlike in SSSP where the overheads were lesser than the computations. Only when the graphs get sufficiently bigger, do the overheads amortize. The EP method, similar to SSSP, consistently delivers better performance than BS. However, its high storage requirements could not be accommodated for the large-sized Graph500 graphs. For the graphs with small diameters, namely the RMAT and ER graphs, the execution time with EP is 48--68\% lesser than that of BS (0.17 MTEPS (BS) vs. 0.54 MTEPS (EP) for RMAT20). For the graphs with large diameters, namely, the road networks, the maximum performance gain with EP over BS is about 10\%.  

Similar to SSSP results, the WD method performs the best among the node-based approaches for graphs with small diameters for the BFS application. The NS method involves considerable overhead for these graphs. For graphs with large diameters, the NS method performs the best since it incurs lesser one-time overheads. In case of relatively larger graphs such as Graph500, HP performs considerably ($>$2$\times$) better than BS, while the EP method fails
to complete execution due to insufficient memory.


\subsection{Performance, Space Complexity and Implementation Tradeoffs}

\begin{figure}
\centering
\includegraphics[width=0.5\linewidth]{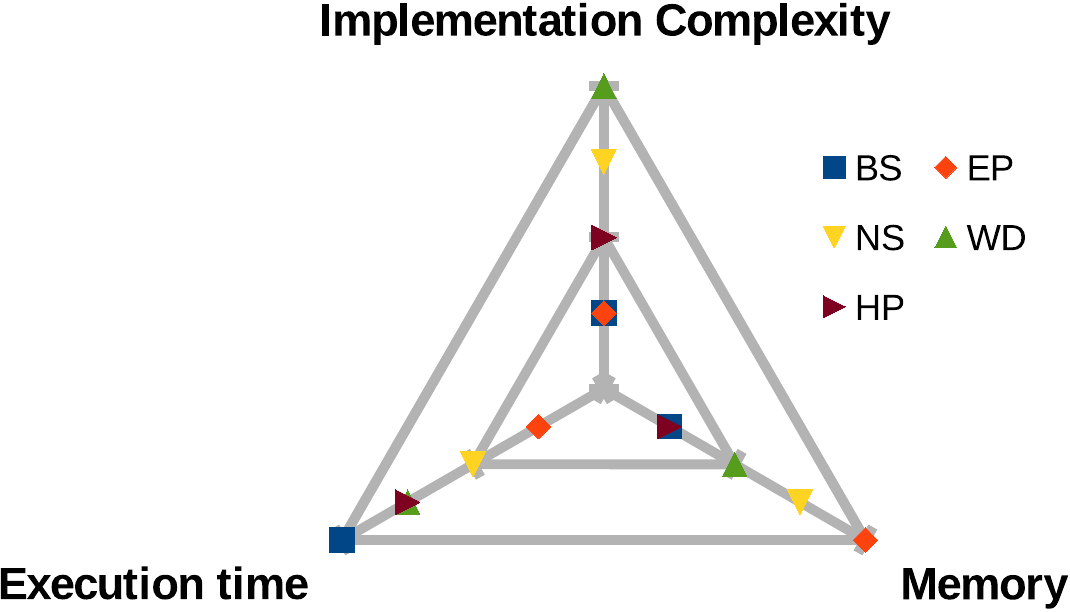}
\caption{Overall comparison of strategies. Each axis represents a ranked order. A strategy closer to the origin (at the center) is ranked higher.}
\label{expt:overall}
\end{figure}

While the previous section focused on the performance aspects of the strategies, in this section we compare the strategies in terms of three axes of comparison: (i) execution time, (ii) memory requirement, and (iii) implementation complexity. The first two are quantitative, while the last one is a qualitative assessment out of our experience. Figure~\ref{expt:overall} shows the relative rankings of the strategies in terms of the three aspects. In each axis, a strategy that is closer to the origin is superior in terms of the corresponding factor.

Overall, it is clear that no one technique fares in all aspects. This suggests that we may have to use different load balancing strategies depending upon the performance requirements, amount of GPU memory and personnel expertise available. Despite the lack of a clear winner, Edge-based Processing (EP) ranks better on two axes (Execution Time and Implementation Complexity). This makes it a more desirable option when the amount of memory is not an issue. 
Node-based processing (which we call as baseline BS) is also easy to implement, and has a low memory requirement (due to CSR representation), but in our experience, performs the worst. Another useful choice could be Hierarchical Processing (HP). It incurs lesser memory penalty and has a moderate implementation complexity. HP does not fare well in terms of performance for small graphs but performs well for large graphs.
Workload Decomposition (WD) and Node Splitting (NS) could be the methods of choice when performance is more important but memory is insufficient to execute EP. NS (implemented as a static phase) is likely to perform better than WD despite graph modification, but incurs larger memory overhead.

\subsection{Degree Distribution due to Node Splitting}
\begin {figure}
\centering
\subfigure[rmat20. MDT=118]{
  \includegraphics[scale=0.2]{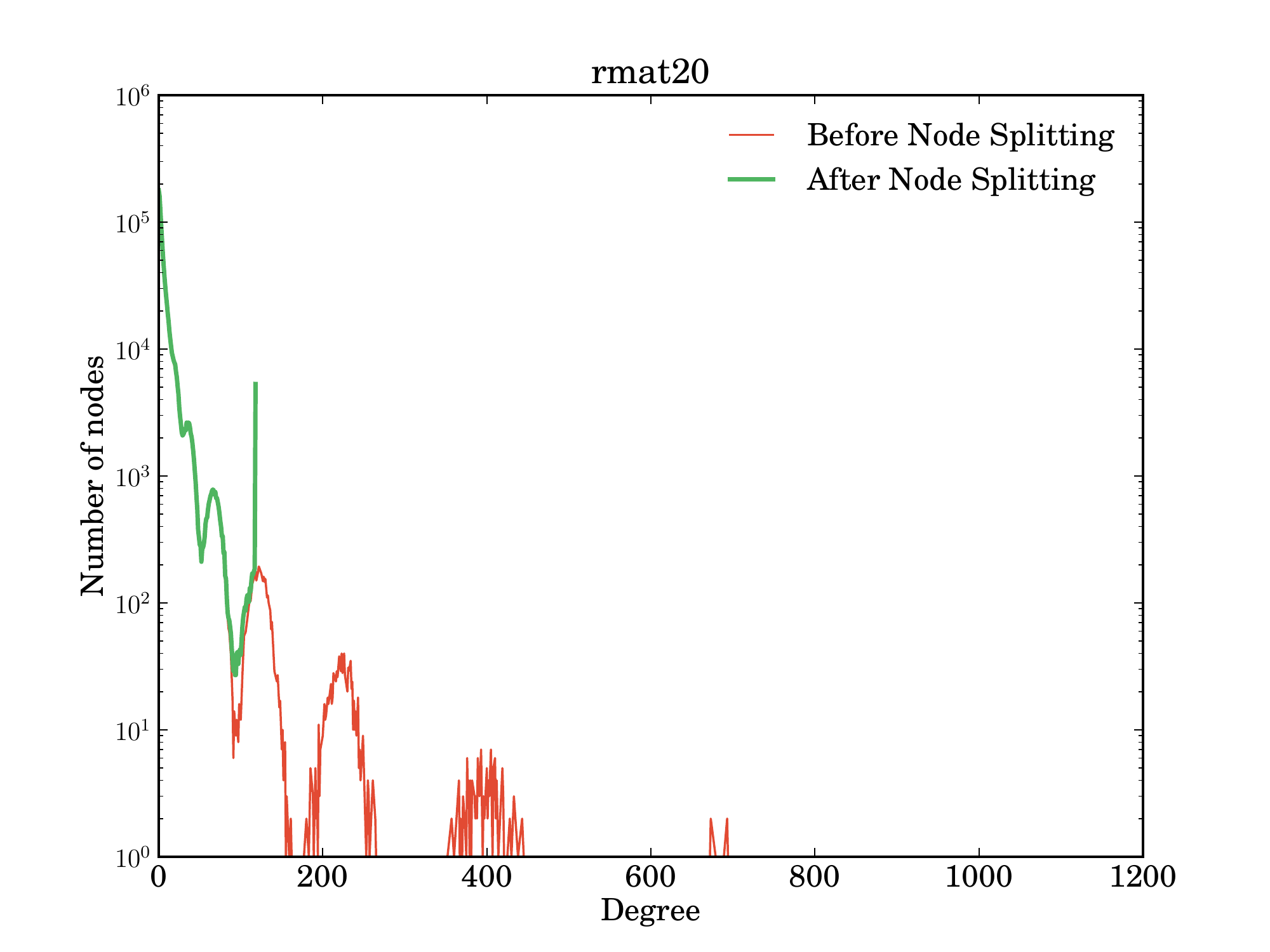}
  \label{rmat20_degNS}
}
\subfigure[ER23. MDT=3]{
  \includegraphics[scale=0.2]{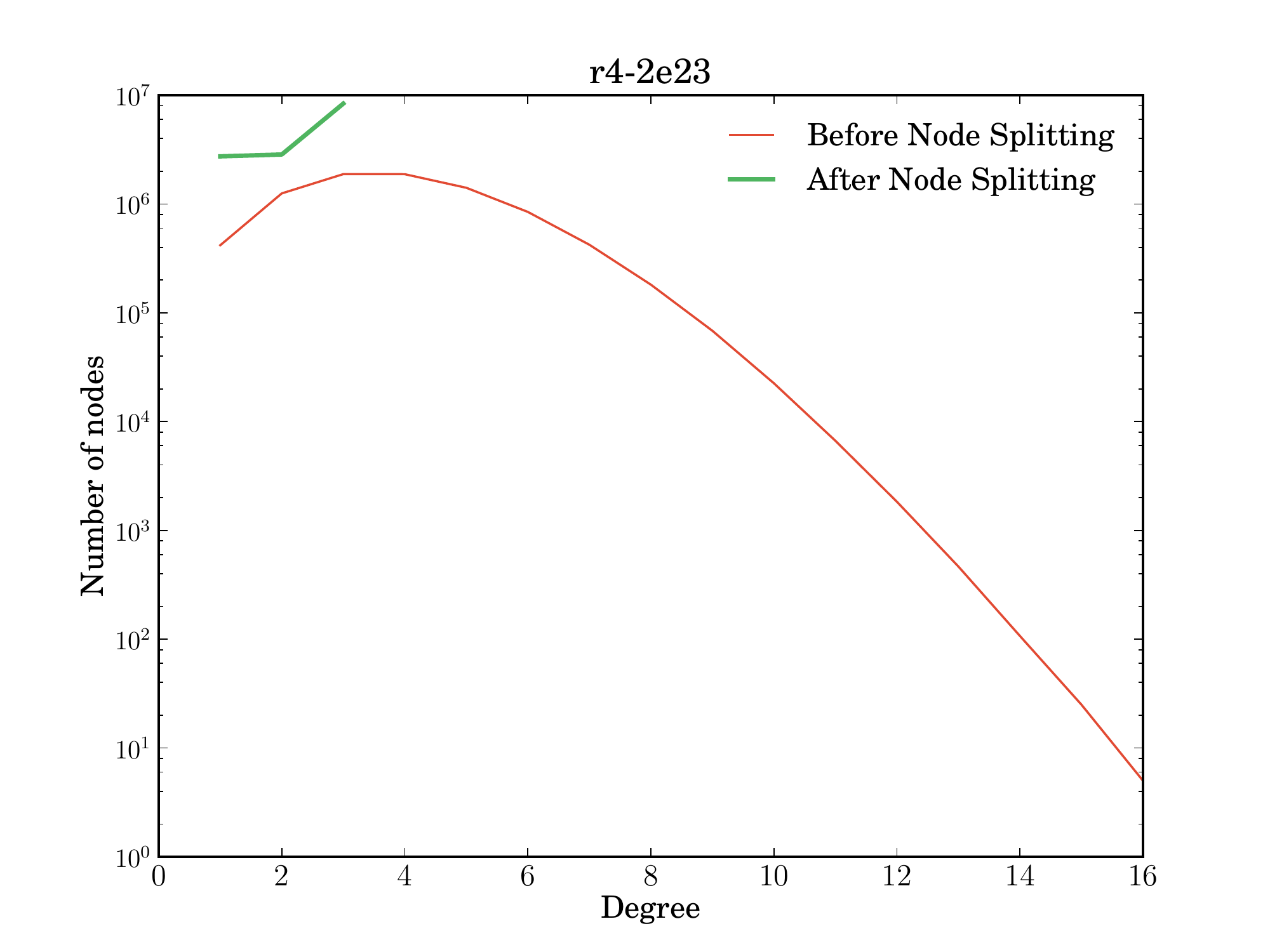}
  \label{r42e23_degNS}
}
\caption{Degree Distributions of Graphs Before and After Node Splitting}
\label{ns-degreedist}
\end{figure}

The NS method modifies the graph by creating child-nodes to distribute the outdegrees.
Therefore, the degree distribution in the modified graph differs from the original.
Figure~\ref{ns-degreedist} shows the distribution of out-degrees of the nodes before (red curve) and after (green curve) node splitting for two synthetic graphs.
The maximum degree thresholds (MDTs) determined using the histogram approach are also shown. 
It is evident from the figure that NS achieves a better load balancing by confining all the nodes to outdegrees within a small range (represented by green curves). We obtained similar results for the other graphs.
It should also be noted that by exploiting histogramming, the MDT does not get biased to a range based on the graph size. 
For instance, for road networks and random graphs, MDT is 2--4 whereas for RMAT graph, it gets rightly computed as 118.

\subsection{Work Chunking Optimization for Edge-based Parallelism}
While using atomic operations to add edges to the worklist in the GPU kernel, we use work chunking in which we collect all edges of a node and add them together using a single atomic operation. 
We compare this strategy with the default strategy of using an atomic operation for adding every edge. 
Figure~\ref{work-chunk-table} shows the speedups obtained due to work chunking EP method over the the default EP method.
We find that work chunking results in 1.11--3.125, with an average of 1.82, speedups over the default method.

\REM {
\begin{table}
\small
\centering
\caption{Benefits of Work Chunking for Reducing Atomic Operations}
\begin{tabular}{|>{\hfill}p{0.55in}|>{\hfill}p{0.45in}|>{\hfill}p{0.5in}|>{\hfill}p{0.55in}|>{\hfill}p{0.55in}|}
\hline 
{\em Graph} & \multicolumn{2}{|c|}{\em BFS} & \multicolumn{2}{|c|}{\em SSSP} \\
\cline{2-5}
& EP (msecs) & EP$_{\mbox{chunk}}$  (msecs) &  EP (msecs) & EP$_{\mbox{chunk}}$ (msecs) \\
\hline \hline
rmat20 & 36.10 & 14.84 & 297.75 & 95.23 \\
\hline
road-FLA & 592.08 & 597.24 & 3535.33 & 2479.11 \\
road-W & 993.44 & 990.82 & 18421.72 & 11790.62 \\
road-USA & 2116.23 & 1865.22 & 228329.00 & 125358.55 \\
\hline
ER20 & 20.47 & 12.62 & 61.94 & 37.19 \\
ER23 & 162.57 & 84.93 & 674.17 & 292.9 \\
\hline 
\end{tabular}
\label{work-chunk-table}
\end{table}
}
\begin{figure}
\centering
  \includegraphics[width=0.70\linewidth]{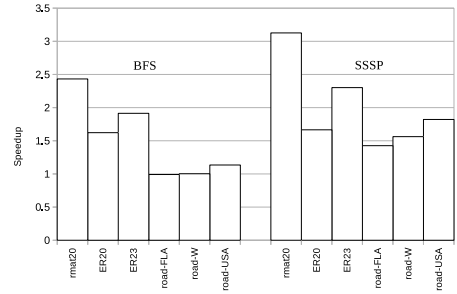}
  \caption{Benefits of Work Chunking in Edge-based Processing}
  \label{work-chunk-table}
\end{figure}

\section{Related Work}
\label{related}

While we have employed fundamental algorithms for BFS and SSSP, various optimized algorithms and implementations for these and other graph applications have been developed on a variety of architectures, including distributed and shared-memory supecomputers, and multi-core machines~\cite{madduri-fasterparallelbc-ipdps2009, kulkarni-lonestar-ispass2009, agarwal-scalablegraphexploration-sc2010, buluc-parallelBFS-sc2011, ediger-graphct-tpds2013,yoo05, bader06a, kulkarni07}. BFS has received significant attention over the years~\cite{Luo:2010:EGI:1837274.1837289,hongBFS,merrill-scalablegputraversal-ppopp2012,gharaibeh12}. The work by Merrill et al.~\cite{merrill-scalablegputraversal-ppopp2012} has developed a work-efficient queue-based algorithm for BFS. $\Delta-$stepping or the derivations of it \cite{meyer-stepping-ja2003, venkatesan-scalablesssp-ipdps2014} are commonly used for SSSP.  Harish and Narayanan~\cite{harish07} describe CUDA implementations of graph algorithms such as BFS and single-source shortest paths computation.  Vineet et al.~\cite{vineet09} and Nobari et al.~\cite{nobari12} propose computing the minimum spanning tree and forest. The primary objective of our work is to propose and evaluate load balancing strategies within a common framework. While we have used the LoneStar-GPU framework and algorithms, our strategies are equally applicable to the above mentioned optimized algorithms as well.

The work by Merrill et al.~\cite{merrill-scalablegputraversal-ppopp2012} has implemented BFS traversal on GPUs using prefix sum computations. The work introduces techniques for local duplication detection to avoid race condition, gathering neighbors for edges, concurrent discovery of duplicates, and strategies for graph contraction and expansion.
Nasre et al. have implemented topology and data-driven versions of several graph applications and have quantitatively compared the two versions~\cite{nasre-datavstoplogy-ipdps2013}. In the topology-driven algorithms, GPU threads are spawned for all nodes in a graph, while in the data-driven algorithms, worklists are built dynamically and threads are spawned corresponding to only active elements/nodes in a time step. In another work, Nasre et al. have developed execution strategies to address challenges related to {\em morph graphs} in which the structure of the graph changes during execution~\cite{nasre-morphgpus-ppopp2013}. They propose optimizations for concurrent node addition, deletion and refinement including reorganizing memory layout, conflict resolution, adaptive parallelism and reducing warp divergence.
The work by Gharaibeh et al.~\cite{gharaibeh-graphsgpus-ipdps2013}  proposed hybrid executions of graph applications utilizing both CPU and GPU cores. The work devises and compares different strategies for partitioning the graph nodes among the CPUs and GPUs.

All these efforts assign the GPU threads to the nodes of the graph, thus performing node-based parallelism. None of these strategies addresses the resulting load imbalance due to node-level parallelism. Sariy{\"u}ce et al.~\cite{sariyuce-bc-gpgpu2013} evaluate both node-based and edge-based parallelism for the betweenness centrality problem. They identify the load imbalance in the node-based parallelism and show that the edge-based parallelism results in good load balance. They also proposed the concept of virtual nodes in which duplicate nodes are created for the actual nodes with high out-degrees. One of our strategies, namely, {\em the node splitting} approach, is similar to the virtual nodes concept. However, in our method, the node-splitting level or the number of virtual nodes is determined automatically using a novel heuristic. Our work is also more comprehensive since it considers multiple load balancing strategies and multiple graph applications. Our work also proposes a novel hierarchical processing method for load balancing. While for the betweenness centrality problem, the authors show that the virtual node strategy performs uniformly better than the edge-based parallelism, we show cases in which the edge-based parallelism gives the best results and analyze the reasons.
We also show that different application scenarios demand different strategies and there is no \textit{one-size-fits-all} solution.

\section{Conclusions and Future Work}
\label{con_fut}

In this paper, we had evaluated four load balancing strategies for BFS and SSSP applications for different graphs. We found that the edge-based processing method performs the best giving about 10\% better performance than the baseline for BFS, and about 60-80\% better performance than the baseline for SSSP. Among the node-based strategies, the workload decomposition method performs the best for graphs with small diameters while the node splitting method performs the best for graphs with large diameters. While the node-based strategies gave worse performance than the baseline in BFS, all our load balancing strategies gave significantly better results (at least 20\% better) than the baseline for SSSP. This shows that load balancing becomes very essential for computationally-intensive graph applications especially for large graphs.
For very large graphs in which some of our load balancing strategies cannot be executed due to memory constraints, our novel hierarchical processing method proposed in this work gives 48-75\% reduction in execution time compared to the baseline.
In future, we plan to explore our strategies for other graph applications including minimum spanning tree and betweenness centrality applications. We also plan to explore dynamic parallelism offered by modern GPU architectures for load balancing graphs. Finally, we plan to build data reorganization strategies for improved coalescing.

\bibliographystyle{IEEEtran}
\bibliography{loadbalancing,others}

\end{document}